\documentclass[11pt,a4paper]{article}

\usepackage{amsmath,amssymb,amsfonts,amscd,amsthm,amsopn,epsfig}
\usepackage{latexsym,verbatim}
\usepackage{mathrsfs}
\usepackage{enumerate}
\usepackage{mathtools}
\usepackage[english]{babel}
\usepackage[numbers,square,sort&compress]{natbib}

\usepackage{color}

\newtheorem{thm}{Theorem}[section]
\newtheorem{prop}{Proposition}[section]

\newtheorem{cor}{Corollary}[section]
\newtheorem{Def}{Definition}[section]

\newtheorem{rmk}{Remark}[section]

\newcommand{\so}{{\scriptscriptstyle \rm I}}
\newcommand{\st}{{\scriptscriptstyle \rm I\hspace{-1pt}I}}
\newcommand{\sth}{{\scriptscriptstyle \rm I\hspace{-1pt}I\hspace{-1pt}I}}
\newcommand{\qo}{{\rm i}}
\newcommand{\qt}{{\rm ii}}
\newcommand{\qth}{{\rm iii}}

\newcommand{\be}[1]{\begin{equation}\label{#1}}
\newcommand{\ba}[1]{\begin{multline}\label{#1}}
\newcommand{\ee}{\end{equation}}
\newcommand{\ea}{\end{eqnarray}}

\newcommand{\cV}{\mathcal{V}}

\newcommand{\bu}{\bar u}
\newcommand{\bv}{\bar v}
\newcommand{\bw}{\bar w}

\newcommand{\bxi}{\bar{\xi}}
\newcommand{\bet}{\bar{\eta}}

\newcommand{\ben}{\begin{eqnarray}}
\newcommand{\een}{\end{eqnarray}}

\newcommand\cH{{\mathcal H}}

\newcommand{\kp}{ \kappa^+}
\newcommand{\km}{ \kappa^-}
\newcommand{\kab}{\tilde \kappa}

\newcommand\CC{\mathbb C}

\usepackage{ulem}

\def\<{\langle}
\def\>{\rangle}
\def\rvec{|0\>}

\def\lvec{\<0|}

 \makeatletter
 \@addtoreset{equation}{section}
 \makeatother

\begin{document}

\vspace{4pt}

\begin{center}
\begin{LARGE}
{\bf Scalar product of twisted XXX modified Bethe vectors }
\end{LARGE}

\vspace{4pt}

\begin{large}
{S. Belliard, N.~A.~Slavnov${}^\ddagger$\ and B.~Vallet${}^\dagger$\  \footnote{samuel.belliard@gmail.com, nslavnov@mi.ras.ru, benoit.vallet@u-psud.fr}}
\end{large}

 \vspace{6mm}

  \vspace{4mm}

${}^\dagger$ {\it Institut de Physique Th\'eorique, DSM, CEA, URA2306 CNRS Saclay, F-91191, Gif-sur-Yvette, France}

 \vspace{4mm}

${}^\ddagger$  {\it Steklov Mathematical Institute of Russian Academy of Sciences, Moscow, Russia}

\end{center}

\vspace{4mm}

\begin{abstract}
We consider closed XXX spin chains with broken total spin $U(1)$ symmetry within the framework of the modified algebraic Bethe ansatz. We study multiple
actions of the modified monodromy matrix entries on the modified Bethe vectors. The obtained formulas of the multiple actions allow us to
calculate the scalar products of the modified Bethe vectors. We find an analog of Izergin--Korepin formula for the scalar products.
This formula involves modified Izergin determinants and can be expressed as sums over partitions of the Bethe  parameters.
\end{abstract}

 \vspace{4mm}


 \vspace{4mm}

%


\section{Introduction}
{ The } recent development of the algebraic Bethe ansatz (ABA) for the models without $U(1)$ symmetry\footnote{{ { Namely models for which} the total spin operator or Cartan operator do not commute with the transfer matrix and the Hamiltonian.}}, the so called {\it modified algebraic Bethe ansatz} (MABA),
gives access to the spectrum and associated eigenstates of the models (see \cite{BP15,BSV18} for the case of the twisted XXX spin chain and references
therein for other models).
A further natural task is to calculate the correlation functions within the framework of this method. Development in this direction { would} allow to adapt
the technique of the { usual} ABA { for the} { study { of} correlation functions for models with $U(1)$ symmetry}  \cite{BogIK93L} to  models without $U(1)$ symmetry. In turn, this { would} allow to obtain
exact solutions in a wide range of fields, such as statistical physics, condensed matter physics, high energy physics, mathematical
physics, and so on.

In the study of correlation functions within the framework of the ABA, 
{ the scalar products of} Bethe { vectors} { play an important role}  \cite{Kor82,BogIK93L,Sla89,Res86,BPRS12a,HutLPRS17,Sla15a}. { Whereas the scalar products are known}, one can compute the
form factors of local operators \cite{KitMT99,MaiT00,KitKMST09,KitKMST11a,BPRS12aa,PakRS14a,PakRS15a,HutLPRS16,FukS17,DugGK13,PozOK12}.
In turn, knowing the form factors, it is possible to calculate the correlation functions by means of their form factor expansion
\cite{KitKMST11b,KitKMST12,KitKMT14,DugGKS17,GohKKKS17,CauM05,CauCS07,PanC14,KR16}.

{ The} calculation of the scalar products is based on the { formulas for the} multiple action \footnote{{ By the multiple action of an operator, we refer to the application of a product of operators of its kind.}} of the monodromy matrix entries on the Bethe vectors
\cite{KitMT00,BPRS12b,HutLPRS16b,Sla16a}.  The specificity of the MABA is that the action of the elements of the monodromy matrix on the
highest weight vector is nonstandard. Usually, this vector is an eigenvector of the diagonal elements and
it is annihilated by the { lower}-triangular part
of the monodromy matrix. However, the monodromy matrix of MABA is obtained from the usual one by means of a non-diagonal twist transformation.
This transformation does not affect the commutation relations between the matrix elements, but changes their actions on the highest weight
vector\footnote{{ A} similar transformation occurs on the framework of the so-called $B^{good}$ operator, see \cite{Bgood,BS18,LS18}.  }.
In particular,  the latter is no longer an eigenvector of the diagonal entries of the monodromy matrix. As a result, the
multiple { actions} formulas change significantly.

In this paper we consider $\mathfrak{gl}_2$-invariant integrable models. An example of such models is the XXX spin-$\frac{1}{2}$ chain with the Hamiltonian
\ben\label{HXXX}
H=\,\sum_{k=1}^N\Big(\sigma^{x}_k\otimes\sigma^{x}_{k+1}+\sigma^{y}_k\otimes\sigma^{y}_{k+1}+\sigma^{z}_k\otimes\sigma^{z}_{k+1}\Big),
\een
subject to the following non-diagonal boundary conditions:
\ben
 \label{cs1}&&\gamma \sigma^{x}_{N+1}=
 \frac{ \kab^2+ \kappa^2-\kappa_+^2-\kappa_-^2}{2}\sigma^{x}_1
 +i\frac{\kappa^2-\kab^2-\kappa_+^2+\kappa_-^2}{2}\sigma^{y}_1
 +(\kappa \kappa_- -\kab \kappa_+)\sigma^{z}_1, \\
 \label{cs2} &&\gamma \sigma^{y}_{N+1}= i\frac{\kab^2- \kappa^2-\kappa_+^2+\kappa_-^2}{2}\sigma^{x}_1
 +\frac{\kab^2+\kappa^2+\kappa_+^2+\kappa_-^2}{2}\sigma^{y}_1
 -i(\tilde \kappa \kappa_++\kappa \kappa_-)\sigma^{z}_1, \\
 \label{cs3}  &&\gamma \sigma^{z}_{N+1}=
(\kappa \kappa_+-\tilde \kappa \kappa_-)\sigma^{x}_1
  +i(\tilde \kappa \kappa_-+\kappa \kappa_+)\sigma^{y}_1
  +(\tilde \kappa \kappa+\kappa_+ \kappa_-)\sigma^{z}_1.
\een
The twist parameters  { $\{\kappa,\tilde \kappa,\kappa_+,\kappa_-\}$ are generic complex numbers} and  $\gamma =\kab\kappa -\kappa_+\kappa_-$. The Pauli matrices\footnote{
 $
\sigma^{z}=\left(\begin{smallmatrix}
       1 & 0\\
      0 & -1      \end{smallmatrix}\right),\quad
\sigma^{+}=\left(\begin{smallmatrix}
       0 & 1\\
      0 & 0      \end{smallmatrix}\right),\quad
\sigma^{-}=\left(\begin{smallmatrix}
       0 & 0\\
      1 & 0      \end{smallmatrix}\right),\quad
      \sigma^{x}=\sigma^{+}+\sigma^{-}, \quad  \sigma^{y}=i(\sigma^{-}-\sigma^{+}) $.}
      $\sigma^{\alpha}_k${ ,} with $\alpha=x,y,z${ ,}  act non-trivially on the $k${ -}th component of the quantum space $\cH= \otimes_{{ j}=1}^N V_{ j}$ with $V_{ j}=\CC^2$. { The twist parameters   $\{\kappa,\tilde \kappa,\kappa_+,\kappa_-\}$  are { the} entries of the most general { $2\times 2$  matrix $K$} that will be introduced in the framework of the ABA
      { (see section~\ref{SS-MAMOD} and also in \cite{BP15,BSV18}).} }

{ Our consideration is not restricted to the Hamiltonian \eqref{HXXX} only. Actually,
we consider { a} more general case with arbitrary highest weight representation and arbitrary non-diagonal twist transformation of the monodromy matrix.} We find the multiple actions of the {\it modified operators} on the  {\it modified Bethe vectors}. { { This} corresponds to the repeated action of the same operator { which depends, in general}{ , on} different parameters.}   This allows us to find a closed expression for the scalar product of two modified Bethe vectors. Multiple action formulas of the usual ABA are expressed in terms of a partition function of the six-vertex model with domain wall boundary condition \cite{Kor82}. { This} latter has { an} explicit representation in terms of the Izergin determinant \cite{Ize87}. Within the
framework of { the} MABA one deals with certain deformation of the Izergin determinant that we call {\it a modified Izergin determinant}. It depends on the parameters of the modified Bethe vectors, but also on the twist parameters. Remarkably, the multiple action formulas and the scalar products of Bethe vectors, being
written in terms of the  modified Izergin determinant{ ,} have almost the same form as their analogs in the usual ABA.

This paper is organized as follows. In section~\ref{MA-IK} we recall the main tools of the ABA. In  section~\ref{A-MDID} we introduce our
notation and the modified Izergin determinant.  We recall multiple actions and a scalar product formula within the standard framework of the ABA in section~\ref{Mul-Ac-Vac}.
In section~\ref{SS-MAMOD} we introduce the modified operators and consider their multiple actions on the modified Bethe vectors. Section~\ref{MSP} is devoted to the calculation of the scalar product of modified Bethe vectors. Auxiliary formulas are gathered in appendices. In appendix~\ref{A-PMID} we { list some} properties of the modified Izergin determinant. In appendix~\ref{ComY} we give { simple} and multiple commutation relations of the monodromy matrix entries
within the standard framework of the ABA. Appendix~\ref{SymYangian} contains a description of a special automorphism of { the} Yangian
of $\mathfrak{gl}_2$.


\section{Basic notions}\label{MA-IK}

{ We consider integrable models described by a rational $R$-matrix belonging to $End(\CC^2\otimes \CC^2)$:}
 \be{Rm}
 R(u)=\frac{u}{c} I+\,P.
 \ee
Here $c$ is a constant, $I=\sum_{i,j=1}^2 E_{ii}\otimes E_{jj}$ is the identity operator
on $\CC^2\otimes \CC^2$, $P=\sum_{i,j=1}^2 E_{ij}\otimes E_{ji}$
is the permutation operator on $\CC^2\otimes \CC^2$, and { are elementary unites:} $(E_{ij})_{kl}=\delta_{ik}\delta_{jl}$.
{ The  $R$-matrix  \eqref{Rm} solves
the Yang--Baxter equation
 \be{YB}
  R_{12}(u-v) R_{13}(u-w) R_{23}(v-w)= R_{23}(v-w) R_{13}(u-w) R_{12}(u-v).
 \ee
This equation holds in the tensor product $V_1 \otimes V_2 \otimes V_3$, where each $V_k\sim\CC^2$.
The $R$-matrix $R_{ij}$ acts nontrivially in the spaces $V_i$ and $V_j$, while it acts as the identity
operator in the remaining space.}
The $R$-matrix \eqref{Rm} is $\mathfrak{gl}_2$-invariant { (and therefore, $GL(2)$-invariant)}, and thus,
\ben\label{twist-inv}
{ [R_{ab}(u),K_a+ K_b]}=[R_{ab}(u),K_a K_b]=0,
\een
for any matrix $K\in End(\CC^2)$.

{ Another important object of the ABA is a monodromy matrix $T(u)$:}
\ben\label{MonoT}
T(u)=\begin{pmatrix} t_{11}(u)& t_{12}(u)\\ t_{21}(u)&t_{22}(u)\end{pmatrix}.
\een
{ The operator-valued entries of the monodromy matrix $t_{ij}(u)$ depend on the complex $u$ (the spectral parameter) and
act in a Hilbert space $\mathcal{H}$ of the associated quantum model.} This matrix  satisfies an RTT relation
\ben\label{RTT}
 R_{ab}(u-v)T_a(u)T_b(v)=T_b(v)T_a(u) R_{ab}(u-v),
\een
which encodes the commutation relations of the entries $t_{ij}(u)$ (see Appendix~\ref{ComY}). These commutation
relations generate a quantum group algebra, so called Yangian of $\mathfrak{gl}_2$.

{ We assume that the Hilbert space $\mathcal{H}$ contains the highest weight vector $\rvec$ possessing
the following properties:}
\ben\label{HWRG}
t_{ii}(u)\rvec=\lambda_i(u)\rvec, \quad t_{21}(u)\rvec=0.
\een
Here $\lambda_i(u)$ are some complex valued functions. { These functions fix a highest weight representation
$\cV(\lambda_1(u), \lambda_2(u))$ of the RTT-algebra \eqref{RTT}. The action of the operator $t_{12}$ on $\rvec$
is free. A state obtained by the successive action of $t_{12}$ on the highest weight vector is called a Bethe vector:}
\ben\label{BVdef}
 t_{12}(\bv)\rvec= \prod_{i=1}^m t_{12}(v_i)\rvec
\een
where $m=0,1,\dots$, and  $\bv=\{v_1,\dots,v_m\}$.

To study scalar products of Bethe vectors we  also use the dual highest weight vector $\lvec$ defined by
\ben\label{dHWRG}
\lvec t_{ii}(u)=\lambda_i(u)\lvec, \quad \lvec t_{12}(u)=0, \quad \lvec 0\>=1.
\een
Here the functions $\lambda_i(u)$ are the same as in \eqref{HWRG}.



\section{Notation and modified Izergin determinant  \label{A-MDID}}

Let us define { the} rational functions
\ben\label{gfh}
g(u,v)=\frac{c}{u-v},\quad f(u,v)=1+g(u,v)=\frac{u-v+c}{u-v}, \quad h(u,v)= \frac{f(u,v)}{g(u,v)}= \frac{u-v+c}{c},
\een
{ where $c$ is the constant entering the $R$-matrix \eqref{Rm}.}
Actually, all these functions depend on the difference of their arguments{ . H}owever we do not stress this dependence.
{ This will { in particular} allow us to use a special shorthand notation (see \eqref{shn}).}
It is easy to see that the functions introduced above possess the following  properties:
\ben\label{x-gfh}
\chi(u,v)\Bigr|_{c\to -c}=\chi(v,u), \qquad \chi(-u,-v)=\chi(v,u),\qquad  \chi(u-c,v)=\chi(u,v+c),
\een
where $\chi$ is any of the three functions. One can also convinces himself that
\ben\label{gfh-prop}
g(u,v-c)=\frac{1}{h(u,v)},\quad h(u,v+c)= \frac{1}{g(u,v)}, \quad f(u,v+c)=\frac{1}{f(v,u)}.
\een

Below we consider sets of complex parameters and denote them by a bar{ . F}or example, $\bu=\{u_1,\dots,u_n\}$. {{ The }notation
$\bu\pm c$ means that $\pm c$ is added to all the arguments of the set $\bu$.} We agree upon
that the notation $\bu_k$ refers to  { the}  set that is complementary { in $\bar u$} to the element $u_k$, that is, $\bu_k=\bu\setminus u_k$.

To make the formulas more compact, we use a shorthand notation for the products of the rational functions \eqref{gfh}, the operators
$t_{kl}(u)$ \eqref{MonoT}, and their vacuum eigenvalues $\lambda_i(u)$ \eqref{HWRG}. Namely, if  { the}  function (operator) depends on a set of variables (similarly to \eqref{BVdef}), then one should take  { the}  product with respect to the corresponding set. For example,
\be{shn}
t_{kl}(\bu)=\prod_{j=1}^n t_{kl}(u_j), \quad \lambda_{i}(\bu)=\prod_{j=1}^n  \lambda_{i}(u_j), \quad f(z,\bu)=\prod_{j=1}^n f(z,u_j), \quad f(\bu_k,u_k)=\prod_{\substack{j=1\\j\ne k}}^nf(u_j,u_k),
\ee
and so on. Note that due to commutativity of the $t_{kl}$-operators the first product in \eqref{shn} is well defined. Notation
$f(\bu,\bv)$ means  { the}  double product over the sets $\bu$ and $\bv$.
By definition any product
over the empty set is equal to $1$. A double product is equal to $1$ if at least one of the sets is empty.

Later we will extend this convention to the products of matrix elements of the twisted monodromy matrix.

\subsection{Modified Izergin determinant\label{SS-MID}}

In many formulas of the ABA the Izergin determinant appears \cite{Kor82,Ize87}. Within the framework
of  { the} MABA we have to deal with a deformation of this object that we call a modified Izergin determinant.

\begin{Def}
Let $\bu=\{u_1,\dots,u_n\}$,  $\bv=\{v_1,\dots,v_m\}$ and $z$ be a complex number.
Then  the modified Izergin determinant $K_{n,m}^{(z)}(\bu|\bv)$
is defined by
\be{defKdef1}
K_{n,m}^{(z)}(\bu|\bv)=\det_m\left(-z\delta_{jk}+\frac{f(\bu,v_j)f(v_j,\bv_j)}{h(v_j,v_k)}\right).
\ee
Alternatively the modified Izergin determinant can be presented as
\be{defKdef2}
K_{n,m}^{(z)}(\bu|\bv)=(1-z)^{m-n}\det_n\left(\delta_{jk}f(u_j,\bv)-z\frac{f(u_j,\bu_j)}{h(u_j,u_k)}\right).
\ee
\end{Def}

The proof of the equivalence of representations \eqref{defKdef1} and \eqref{defKdef2} can be found { in proposition~4.1 of} \cite{GorZZ14}.
It is based on the recursive property \eqref{resK}. The modified Izergin determinant is related to the partial domain wall partition functions \cite{FW12}.
{ Other correspondences will be discussed elsewhere.}

It is  also convenient to introduce a conjugated modified  Izergin determinant as
\be{CdefKdef1}
\overline{K}_{n,m}^{(z)}(\bu|\bv)=K_{n,m}^{(z)}(\bu|\bv)\Bigr|_{c\to -c}
=\det_m\left(-z\delta_{jk}+\frac{f(v_j,\bu)f(\bv_j,v_j)}{h(v_k,v_j)}\right),
\ee
or equivalently
\be{CdefKdef2}
\overline{K}_{n,m}^{(z)}(\bu|\bv)=(1-z)^{m-n}\det_n\left(\delta_{jk}f(\bv,u_j)-z\frac{f(\bu_j,u_j)}{h(u_k,u_j)}\right).
\ee

In the particular case $z=1$ and $\#\bu=\#\bv=n$ the modified Izergin determinant
turns into the ordinary Izergin determinant, that we traditionally denote by $K_{n}(\bu|\bv)$:
\be{ModI-OrdI}
K_{n,n}^{(1)}(\bu|\bv)=K_{n}(\bu|\bv).
\ee
This property can be seen from the recursion \eqref{resK} and the initial condition \eqref{K1}. It also follows
from \eqref{defKdef2} that
\be{K10}
K_{n,m}^{(1)}(\bu|\bv)=0,\qquad \text{for} \qquad n<m.
\ee
Other properties of the modified Izergin determinant are collected in Appendix~\ref{A-PMID}.

\section{Multiple actions\label{Mul-Ac-Vac}}


Actions of the operators $t_{ij}(u)$ on the Bethe vectors \eqref{BVdef} were computed in \cite{FadST79} (see also \cite{BogIK93L}).
To study the problem of the scalar products one should calculate multiple actions of the form
\be{MABVdef}
t_{ij}(\bu) t_{12}(\bv)\rvec.
\ee
Here, according to the convention on the shorthand notation \eqref{shn} $t_{ij}(\bu)$ is the product of the operators
$t_{ij}$ over  { the}  set $\bu=\{u_1,\dots,u_n\}$.
Such multiple actions  of the monodromy matrix entries were found in \cite{BPRS12b} for the models with $\mathfrak{gl}_3$-invariant
$R$-matrix. In { our} particular case, these formulas give the multiple actions for models with
$\mathfrak{gl}_2$-invariant $R$-matrix.

Multiple action formulas are given in terms of sums over partitions of the set $\bw=\{\bu,\bv\}$ into subsets. Here and below
we mostly denote the subsets by Roman subscripts (except for some special cases). Notation $\bw \Rightarrow \{\bw_{\so},\bw_{\st}\}$
(and similar ones) means that the set $\bw$ is divided into subsets $\bw_{\so}$ and $\bw_{\st}$ such that $\bw_{\so}\cup\bw_{\st}=\bw$ and
$\bw_{\so}\cap\bw_{\st}=\emptyset$.

\begin{prop}\cite{BPRS12b}
Let $\#\bu=n$, $\#\bv=m$,  $\bw=\{\bu,\bv\}$, and $K_{n}$ be the Izergin determinant \eqref{ModI-OrdI}. Then
\ben
&&t_{12}(\bu)t_{12}(\bv)\rvec=t_{12}(\bw)\rvec.
\een
The actions of the diagonal elements $t_{ii}$ are given by
\ben\label{MA1122}
&&t_{11}(\bu)t_{12}(\bv)\rvec= (-1)^n\sum_{\substack{\bw \Rightarrow \{\bw_{\so},\bw_{\st}\}\\ \#\bw_{\so}=n} }
\lambda_1(\bw_{\so})\overline{K}_{n}(\bu|\bw_{\so}-c)f(\bw_{\st},\bw_{\so})t_{12}(\bw_{\st})\rvec,\\
&&t_{22}(\bu)t_{12}(\bv)\rvec=(-1)^n\sum_{\substack{\bw \Rightarrow \{\bw_{\so},\bw_{\st}\}\\ \#\bw_{\so}=n} }
 \lambda_2(\bw_{\so})K_{n}(\bu|\bw_{\so}+c)f(\bw_{\so},\bw_{\st})t_{12}(\bw_{\st})\rvec,
\een
where the sums are taken over partitions { ?$\{ \bu ,\bv \}$=}$\bw \Rightarrow \{\bw_{\so},\bw_{\st}\}$ such that $\#\bw_{\so}=n$.
The action of the elements $t_{21}$ reads
\ben\label{MAt21}
&&t_{21}(\bu)t_{12}(\bv)\rvec=\sum_{\substack{\bw \Rightarrow \{\bw_{\so},\bw_{\st}\}\\ \#\bw_{\so}=\#\bw_{\st}=n} }
 \lambda_2(\bw_{\so}) \lambda_1(\bw_{\st})
K_{n}(\bu|\bw_{\so}+c)\overline{K}_{n}(\bu|\bw_{\st}-c)\nonumber\\
&&\qquad\qquad\qquad\qquad \qquad\qquad \times f(\bw_{\so},\bw_{\st})f(\bw_{\so},\bw_{\sth})f(\bw_{\sth},\bw_{\st})t_{12}(\bw_{\sth})\rvec,
\een
where the sum is taken over partitions $\bw \Rightarrow \{\bw_{\so},\bw_{\st},\bw_{\sth}\}$ such that $\#\bw_{\so}=\#\bw_{\st}=n$.
\end{prop}

Note that the action formulas \eqref{MA1122} are  { the}  direct consequence of the commutation relations \eqref{MCR1122}.

Equation \eqref{MAt21} gives immediate access to the scalar product of Bethe vectors defined by
\be{SP-defin}
S^n_t(\bu,\bv)=\lvec t_{21}(\bu)t_{12}(\bv)\rvec,
\ee
where $\#\bu=\#\bv=n$.

\begin{thm}\label{SC}
Let $\#\bu=\#\bv=n$. Then  the scalar product of two Bethe vectors is given by
\ben\label{SCe}
S^n_t(\bu,\bv)=\sum_{\substack{\bw \Rightarrow \{\bw_{\so},\bw_{\st}\}\\ \#\bw_{\so}=\#\bw_{\st}=n} }
\lambda_2(\bw_{\so}) \lambda_1(\bw_{\st})K_{n}(\bu|\bw_{\so}+c)
\overline{K}_{n}(\bu|\bw_{\st}-c)f(\bw_{\so},\bw_{\st}).
\een
where the sum is taken over partitions $\bw \Rightarrow \{\bw_{\so},\bw_{\st}\}$ such that $\#\bw_{\so}=\#\bw_{\st}=n$.
\end{thm}
{
The sum \eqref{SCe}  can also be written in the form of  { the}  sum over independent partitions of the sets $\bu$ and $\bv$.
Then it corresponds to the  Izergin--Korepin formula \cite{BogIK93L}.
\begin{cor}\label{SCb}
Let $\#\bu=\#\bv=n$. Then  the scalar product of two Bethe vectors is given by
\begin{equation}\label{SCbe}
S^n_t(\bu,\bv)=\sum_{\substack{\bu \Rightarrow \{\bu_{\so},\bu_{\st}\}\\ \bv \Rightarrow \{\bv_{\so},\bv_{\st}\}\\ \#\bu_{\so}=\#\bv_{\so}} }
\lambda_2(\bu_{\so}) \lambda_2(\bv_{\st}) \lambda_1(\bu_{\st}) \lambda_1(\bv_{\so})
K_{n_2}(\bv_{\st}|\bu_{\st})
\overline{K}_{n_1}(\bv_{\so}|\bu_{\so})
f(\bu_{\so},\bu_{\st})f(\bv_{\st},\bv_{\so}),
\end{equation}
where the sum is taken over partitions $\bu \Rightarrow \{\bu_{\so},\bu_{\st}\}$ and $ \bv \Rightarrow \{\bv_{\so},\bv_{\st}\}$ such that $\#\bu_{\so}=\#\bv_{\so}=n_1$, $\#\bu_{\st}=\#\bv_{\st}=n_2$, where $n_1=0,1,\dots,n$ and $n=n_1+n_2$.
\end{cor}
\proof
We set in  \eqref{SCe} $\bw_{\so}\Rightarrow \{\bu_{\so},\bv_{\st}\}$ and $\bw_{\st}\Rightarrow \{\bu_{\st},\bv_{\so}\}$. Let  $\#\bu_{\so}=\#\bv_{\so}=n_1$, $\#\bu_{\st}=\#\bv_{\st}=n_2$, where $n_1=0,1,\dots,n$ and $n=n_1+n_2$. Using \eqref{Kz} and \eqref{oKz} we obtain
\begin{multline}\label{0SCbe}
S^n_t(\bu,\bv)=(-1)^n\sum_{\substack{\bu \Rightarrow \{\bu_{\so},\bu_{\st}\}\\ \bv \Rightarrow \{\bv_{\so},\bv_{\st}\}\\ \#\bu_{\so}=\#\bv_{\so}} }
\lambda_2(\bu_{\so}) \lambda_2(\bv_{\st}) \lambda_1(\bu_{\st}) \lambda_1(\bv_{\so})
K_{n_2}(\bu_{\st}|\bv_{\st}+c)\overline{K}_{n_1}(\bu_{\so}|\bv_{\so}-c) \\
\times
f(\bu_{\so},\bu_{\st})f(\bv_{\st},\bv_{\so})f(\bu_{\so},\bv_{\so})f(\bv_{\st},\bu_{\st}).
\end{multline}
Then the use of   \eqref{Kinv1} and  \eqref{cKinv1} immediately leads us to \eqref{SCbe}.
\endproof
}
\section{Multiple actions of modified operators on Bethe vectors\label{SS-MAMOD}}

A monodromy matrix of MABA is constructed as a twist transformation of the original monodromy matrix \eqref{MonoT}. { In \cite{BP15,BSV18} we discussed the factorisation of the twist matrix $K=BDA$ { (where $D$ is a diagonal matrix), which allows us} to
use the MABA. It { includes} some freedom by the transformation $A \to SA$ and $B\to BS^{-1}$ for any invertible
diagonal matrix { $S$}.
Let us consider the { following} parametrization of the two matrices $A$ and $B$:}
\ben\label{Mat-Tf}
A=\sqrt\mu\Big(\begin{array}{ll} 1&\frac{\rho_2}{\km}\\ \frac{\rho_1}{\kp}&1\end{array}\Big),\quad B=\sqrt\mu\Big(\begin{array}{ll} 1&\frac{\rho_1}{\km}\\ \frac{\rho_2}{\kp}&1\end{array}\Big),\quad \mu=\frac{1}{1-\frac{\rho_1\rho_2}{\kp\km}}.
\een
Here $\rho_i$ and $\kappa^\pm$ are generic parameters.
Due to the property \eqref{twist-inv}, the transformation of the monodromy matrix
\ben\label{MonoV}
\bar T(u)=AT(u)B=\begin{pmatrix} \nu_{11}(u)& \nu_{12}(u)\\ \nu_{21}(u)&\nu_{22}(u)\end{pmatrix}
\een
is an automorphism of the Yangian of $\mathfrak{gl}_2$, i.e. new operators $\nu_{ij}$ satisfy the same commutation relations { as} the $t_{ij}(z)${ ,}
given in Appendix~\ref{ComY}.
However, the actions of the modified operators $\{\nu_{ii}(u),\nu_{21}(u)\}$
on the highest weight vector \eqref{HWRG}  change. It is easy to see that now they are given by
\ben
&&\nu_{11}(u)\rvec= \lambda_1(u)\rvec+\beta_2\nu_{12}(u)\rvec, \label{act-sing-1}\\
&&\nu_{22}(u)\rvec= \lambda_2(u)\rvec+\beta_1\nu_{12}(u)\rvec,\label{act-sing-2}\\
&&\nu_{21}(u)\rvec=\Big(\beta_1\lambda_1(u)+\beta_2\lambda_2(u)\Big)\rvec+\beta_1\beta_2\nu_{12}(u)\rvec,\label{act-sing-3}
\een
where $\beta_i=\frac{\rho_i}{\kp}$.

The modified Bethe vectors are given by
\ben
\nu_{12}(\bar v)\rvec= \prod_{i=1}^m \nu_{12}(v_i)\rvec
\een
with $m=0,1,\dots$. 
Here we extended the convention on the shorthand notation \eqref{shn} to the products of the operators
$\nu_{ij}$. Since the commutation relations of $\nu_{ij}$ are the same as { the} ones of $t_{ij}$, we have, in particular,
$[\nu_{ij}(u),\nu_{ij}(v)]=0$. Thus, the products $\nu_{ij}(\bar v)$ are well defined.

\subsection{Multiple actions of the modified diagonal operators \label{Mul-Ac-Vac1}}

It is clear that changing the action on the highest weight vector leads to a modification of the multiple action formulas.

\begin{prop}\label{mulii}
The multiple actions of the products of { the}  diagonal modified operators $\nu_{ii}(\bar u)$, with $\bar u=\{u_1,...,u_n\}$,
on the modified Bethe vector $\nu_{12}(\bv)|0\rangle$, with $\bar v=\{v_1,...,v_m\}$, are given by
\begin{align}\label{nvac1112}
\nu_{11}(\bu)\nu_{12}(\bv)|0\rangle &=\beta_2^n\sum_{\bw\Rightarrow\{\bw_{\so},\bw_{\st}\}} (-\beta_2)^{-l}
\lambda_1(\bw_{\so}) \overline{K}^{(1)}_{n,l}(\bu|\bw_{\so}-c)
f(\bw_{\st},\bw_{\so})\nu_{12}(\bw_{\st})|0\rangle,\\
\label{nvac2212}
\nu_{22}(\bu)\nu_{12}(\bv)|0\rangle &=\beta_1^n\sum_{\bw\Rightarrow\{\bw_{\so},\bw_{\st}\}}(-\beta_1)^{-l}
\lambda_2(\bw_{\so}) K^{(1)}_{n,l}(\bu|\bw_{\so}+c)
f(\bw_{\so},\bw_{\st})\nu_{12}(\bw_{\st})|0\rangle.
\end{align}
Here $l=\#\bw_{\so}$. The sum is taken over all partitions $\{\bu,\bv\}=\bw\Rightarrow\{\bw_{\so},\bw_{\st}\}$. There is no restrictions on the cardinalities
of the subsets.
The function $K^{(1)}_{n,l}$ and $\overline{K}^{(1)}_{n,l}$  respectively are the modified
Izergin determinants \eqref{defKdef1} and \eqref{CdefKdef1} at $z=1$.
\end{prop}

\begin{rmk}
 The main difference between modified action formulas  and equations \eqref{MA1122} is the replacement of the
ordinary Izergin determinants  with the modified Izergin determinants. This leads to the fact that
there is no restriction on the cardinalities of the subsets in formulas \eqref{nvac1112}, \eqref{nvac2212}.
However, due to the property $K^{(1)}_{n,l}(\bu|\bv)=\overline K^{(1)}_{n,l}(\bu|\bv)=0$ for $n<l$, the summation in
\eqref{nvac1112}, \eqref{nvac2212} is carried out only over those partitions for which $l\le n$.
\end{rmk}

\proof
We give a detailed proof of formula \eqref{nvac1112}. The proof of formula \eqref{nvac2212} is completely analogous.
It also follows form \eqref{nvac1112} due to the symmetry of the Yangian described in appendix~\ref{SymYangian}.

We first consider the case $n=\#\bu=1$. In fact, in this case, equation \eqref{nvac1112} was firstly conjectured in \cite{BP15} and then proved
in \cite{BS18}. Therefore, we consider this case for the sake of completeness only.

Since the operators $\nu_{ij}$ possess the same commutation relations as $t_{ij}$, we
can use \eqref{MCR1122} for $n=1$:
\be{CR1112}
\nu_{11}(u)\nu_{12}(\bv)= -\sum_{\substack{\bw\Rightarrow\{\bw_{\so},\bw_{\st}\}\\ \#\bw_{\so}=1}} \overline{K}_{1}(u|\bw_{\so}-c)f(\bw_{\st},\bw_{\so})\nu_{12}(\bw_{\st})\nu_{11}(\bw_{\so}).
\ee
Here $\bw=\{u,\bv\}$. The sum is taken over partitions $\bw\Rightarrow\{\bw_{\so},\bw_{\st}\}$ such that $\#\bw_{\so}=1$.
Applying this equation to $\rvec$ and using \eqref{act-sing-1} we obtain
\be{SinA1112}
\nu_{11}(u)\nu_{12}(\bv)\rvec= -\sum_{\substack{\bw\Rightarrow\{\bw_{\so},\bw_{\st}\}\\ \#\bw_{\so}=1}}
\overline{K}_{1}(u|\bw_{\so}-c)f(\bw_{\st},\bw_{\so})\nu_{12}(\bw_{\st})
\Bigl(\lambda_1(\bw_{\so})+\beta_2\nu_{12}(\bw_{\so})\Bigr)\rvec.
\ee
The sum over partitions in the term proportional to $\beta_2$ can be computed explicitly. Indeed, we have
$$
-\overline{K}_{1}(u|\bw_{\so}-c)=\frac{c}{u-\bw_{\so}+c}=\frac1{h(u,\bw_{\so})}.
$$
Then
\be{sumtriv}
-\beta_2\sum_{\substack{\bw\Rightarrow\{\bw_{\so},\bw_{\st}\}\\ \#\bw_{\so}=1}}
\overline{K}_{1}(u|\bw_{\so}-c)f(\bw_{\st},\bw_{\so})\nu_{12}(\bw_{\st})
\nu_{12}(\bw_{\so})\rvec=
\beta_2\nu_{12}(\bw)\rvec\;G,
\ee
where
\be{G1}
G=\sum_{\substack{\bw\Rightarrow\{\bw_{\so},\bw_{\st}\}\\ \#\bw_{\so}=1}}  \frac{f(\bw_{\st},\bw_{\so})}{h(u,\bw_{\so})}.
\ee
To calculate the sum over partitions \eqref{G1} it is enough to present it as a contour integral
\be{CI-01}
G=\frac{-1}{2\pi i c}\oint_{ |z|=R\to\infty}\frac{f(\bw,z)}{h(u,z)}\,dz.
\ee
Taking the residue at infinity we obtain\footnote{Recall that $u\in\bw$, and thus, there is no
pole at $z=u+c$.} $G=1$ . Thus,
\be{SinA1112-f1}
\nu_{11}(u)\nu_{12}(\bv)\rvec=\beta_2\nu_{12}(\bw)\rvec -\sum_{\substack{\bw\Rightarrow\{\bw_{\so},\bw_{\st}\}\\ \#\bw_{\so}=1}}
\lambda_1(\bw_{\so})\overline{K}_{1}(u|\bw_{\so}-c)f(\bw_{\st},\bw_{\so})\nu_{12}(\bw_{\st})\rvec.
\ee

It remains to compare the result obtained with equation \eqref{nvac1112} for $n=1$. In this case
either $l=0$ or $l=1$. It is easy to see that the first term in \eqref{SinA1112-f1} corresponds
to the case $l=0$, while the second term gives the sum over partitions for $l=1$.
Thus, the action \eqref{nvac1112} is proved for $n=1$.

To proceed  further we use induction over $n$. Assume that \eqref{nvac1112} holds for some $n-1$. Then the action
of $\nu_{11}(\bu)$ on the modified Bethe vector $\nu_{12}(\bv)|0\rangle$ can be computed as the successive action
of $\nu_{11}(\bu_n)$ and $\nu_{11}(u_n)$ (recall that $\bu_n=\bu\setminus u_n$). At the first step we have
\be{indstep-1}
\nu_{11}(\bu)\nu_{12}(\bv)|0\rangle =\nu_{11}(u_n)\beta_2^{n-1}\sum_{\bxi\Rightarrow\{\bxi_{\so},\bxi_{\st}\}} (-\beta_2)^{-l_{\so}}
\lambda_1(\bxi_{\so}) \overline{K}^{(1)}_{n-1,l}(\bu_n|\bxi_{\so}-c)
f(\bxi_{\st},\bxi_{\so})\nu_{12}(\bxi_{\st})|0\rangle.
\ee
Here $\bxi=\{\bu_n,\bv\}$. The sum is taken over partitions $\bxi\Rightarrow\{\bxi_{\so},\bxi_{\st}\}$, and $l_{\so}=\#\bxi_{\so}$.
Acting with $\nu_{11}(u_n)$ on the vector $\nu_{12}(\bxi_{\st})|0\rangle$ we obtain
\begin{multline}\label{indstep-2}
\nu_{11}(\bu)\nu_{12}(\bv)|0\rangle =\beta_2^{n}\sum_{\bxi\Rightarrow\{\bxi_{\so},\bxi_{\st}\}} (-\beta_2)^{-l_{\so}}
\lambda_1(\bxi_{\so}) \overline{K}^{(1)}_{n-1,l_{\so}}(\bu_n|\bxi_{\so}-c)
f(\bxi_{\st},\bxi_{\so})\\
\times\sum_{\bet\Rightarrow\{\bet_{\so},\bet_{\st}\}} (-\beta_2)^{-k_{\so}}
\lambda_1(\bet_{\so}) \overline{K}^{(1)}_{1,k_{\so}}(u_n|\bet_{\so}-c)
f(\bet_{\st},\bet_{\so})\nu_{12}(\bet_{\st})|0\rangle.
\end{multline}
Here we have one more sum over partitions of the set $\bet=\{u_n,\bxi_{\st}\}\Rightarrow\{\bet_{\so},\bet_{\st}\}$, and $k_{\so}=\#\bet_{\so}$.

Thus, in \eqref{indstep-2}, the set $\{\bu,\bv\}$ eventually is divided into three subsets $\bxi_{\so}$, $\bet_{\so}$, and
$\bet_{\st}$. The subset $\bxi_{\st}$ plays { an} intermediate role and should be understood as
$\bxi_{\st}=\{\bet_{\so},\bet_{\st}\}\setminus\{u_n\}$. The only restriction on these partitions is that $u_n\notin\bxi_{\so}$.

Let $\bw=\{\bu,\bv\}$. Denote $\bxi_{\so}=\bw_{\so}$, $\bet_{\so}=\bw_{\st}$, and
$\bet_{\st}=\bw_{\sth}$. Then $\bxi_{\st}=\{\bw_{\st},\bw_{\sth}\}\setminus\{u_n\}$ and
\be{fff}
f(\bxi_{\st},\bxi_{\so})=\frac{f(\bw_{\st},\bw_{\so})f(\bw_{\sth},\bw_{\so})}{f(u_n,\bw_{\so})}.
\ee
Observe that the { right hand side} of \eqref{fff} vanishes as soon as $u_n\in\bw_{\so}$. Thus, the condition $u_n\notin\bxi_{\so}$ holds automatically.
Equation \eqref{indstep-2} then takes the following form:
\begin{multline}\label{indstep-3}
\nu_{11}(\bu)\nu_{12}(\bv)|0\rangle =\beta_2^{n}\sum_{\bw\Rightarrow\{\bw_{\so},\bw_{\st},\bw_{\sth}\}} (-\beta_2)^{-r_{\so}-r_{\st}}
\lambda_1(\bw_{\so})\lambda_1(\bw_{\st})
\\
\times \overline{K}^{(1)}_{n-1,r_{\so}}(\bu_n|\bw_{\so}-c) \overline{K}^{(1)}_{1,r_{\st}}(u_n|\bw_{\st}-c)
\frac{f(\bw_{\st},\bw_{\so})f(\bw_{\sth},\bw_{\so})f(\bw_{\sth},\bw_{\st})}{f(u_n,\bw_{\so})}
\nu_{12}(\bw_{\sth})|0\rangle.
\end{multline}
Here $r_{\so}=\#\bw_{\so}$ and  $r_{\st}=\#\bw_{\st}$. Let $\{\bw_{\so},\bw_{\st}\}=\bw_0$ and $r_0=\#\bw_0$. Then, we recast \eqref{indstep-3} as follows:
\begin{multline}\label{indstep-4}
\nu_{11}(\bu)\nu_{12}(\bv)|0\rangle =\beta_2^{n}\sum_{\bw\Rightarrow\{\bw_{0},\bw_{\sth}\}} (-\beta_2)^{-r_{0}}
\lambda_1(\bw_{0})f(\bw_{\sth},\bw_{0})\nu_{12}(\bw_{\sth})|0\rangle
\\
\times \sum_{\bw_0\Rightarrow\{\bw_{\so},\bw_{\st}\}}
\overline{K}^{(1)}_{n-1,r_{\so}}(\bu_n|\bw_{\so}-c) \overline{K}^{(1)}_{1,r_{\st}}(u_n|\bw_{\st}-c)
\frac{f(\bw_{\st},\bw_{\so})}{f(u_n,\bw_{\so})}.
\end{multline}
The sum over partitions is now organized in two steps. First, the set $\bw$ is divided into two subsets $\bw_0\Rightarrow\{\bw_{\so},\bw_{\st}\}$.
Then the subset $\bw_0$ is divided once more as $\bw_0\Rightarrow\{\bw_{\so},\bw_{\st}\}$. It is easy to see that the sum over partitions in
the second line of \eqref{indstep-4} reduces to the modified Izergin determinant due to \eqref{CML-2}:
\begin{equation}\label{UseCML-2}
 \sum_{\bw_0\Rightarrow\{\bw_{\so},\bw_{\st}\}}
\overline{K}^{(1)}_{n-1,r_{\so}}(\bu_n|\bw_{\so}-c) \overline{K}^{(1)}_{1,r_{\st}}(u_n|\bw_{\st}-c)
\frac{f(\bw_{\st},\bw_{\so})}{f(u_n,\bw_{\so})}=\overline{K}^{(1)}_{n,r_{0}}(\bu|\bw_{0}-c).
\end{equation}
Thus, we arrive at
\begin{equation}\label{indstep-5}
\nu_{11}(\bu)\nu_{12}(\bv)|0\rangle =\beta_2^{n}\sum_{\bw\Rightarrow\{\bw_{0},\bw_{\sth}\}} (-\beta_2)^{-r_{0}}
\lambda_1(\bw_{0})\overline{K}^{(1)}_{n,r_{0}}(\bu|\bw_{0}-c)f(\bw_{\sth},\bw_{0})\nu_{12}(\bw_{\sth})|0\rangle.
\end{equation}
This equation coincides with \eqref{nvac1112} for $\#\bu=n$ up to the labels of the subsets. \endproof


\subsection{Multiple action of the modified operator $\nu_{21}$\label{Mul-Ac-BV}}

\begin{prop}\label{mul21}
The multiple action of the product of  modified operators $\nu_{21}(\bar u)$, with $\bar u=\{u_1,...,u_n\}$,
on the modified Bethe vector $\nu_{12}(\bv)|0\rangle$, with $\bar v=\{v_1,...,v_m\}$, is given by
\begin{multline}\label{nvac2112}
 \nu_{21}(\bu)\nu_{12}(\bv)|0\rangle=\sum_{\bw\Rightarrow\{\bw_{\so},\bw_{\st},\bw_{\sth}\} }
 (-\beta_1)^{n-l_{\so}}  (-\beta_2)^{n-l_{\st}}\lambda_2(\bw_{\so})\lambda_1(\bw_{\st})
\\
\times  K^{(1)}_{n,l_{\so}}(\bu|\bw_{\so}+c) \overline{K}^{(1)}_{n,l_{\st}}(\bu|\bw_{\st}-c)
f(\bw_{\so},\bw_{\st})f(\bw_{\so},\bw_{\sth})f(\bw_{\sth},\bw_{\st})\nu_{12}(\bw_{\sth})|0\rangle.
\end{multline}
Here $l_{\so}=\bw_{\so}$ and $l_{\st}=\bw_{\st}$. The sum is taken over all partitions
$\{\bu,\bv\}=\bw\Rightarrow\{\bw_{\so},\bw_{\st},\bw_{\sth}\}$.
The function $K^{(1)}_{n,l_{\so}}$ and $\overline{K}^{(1)}_{n,l_{\st}}$  respectively are the modified
Izergin determinants \eqref{defKdef1} and \eqref{CdefKdef1} at $z=1$.
\end{prop}
\proof
To prove \eqref{nvac2112} we first use induction over $m=\#\bv$ and then over $n=\#\bu$.

Let $n=1$ and, hence, $\bu=u$. Note that in spite of the sum in \eqref{nvac2112} is taken over all possible partitions of the set $\bw=\{u,\bv\}$,
in fact, it is restricted by the condition $l_i\le n$ ($i=\so,\st$), because otherwise the modified Izergin determinants vanish. Thus, for $n=1$
the cardinalities of the subsets $\bw_{\so}$ and $\bw_{\st}$ are either $0$ or $1$.
Then, it is easy to see that for $n=1$ and $m=0$, equation \eqref{nvac2112} coincides with the  action formula \eqref{act-sing-3}.

Assume that \eqref{nvac2112} holds for some $m-1$, where $m>0$. Using commutation relation \eqref{comsl2112} we obtain
\be{sing-com}
\nu_{21}(u)\nu_{12}(\bv)|0\rangle=\bigl[\nu_{12}(v_m)\nu_{21}(u)+ g(u,v_m)\big(\nu_{11}(v_m)\nu_{22}(u)
-\nu_{11}(u)\nu_{22}(v_m)\big)\bigr]\nu_{12}(\bv_m)|0\rangle.
\ee
Let us first consider the contribution of the  term $\nu_{12}(v_m)\nu_{21}(u)$. Due to the induction assumption we have
\begin{multline}\label{1-contr1}
 \nu_{12}(v_m)\nu_{21}(u)\nu_{12}(\bv_m)|0\rangle=\sum_{\bxi\Rightarrow\{\bxi_{\so},\bxi_{\st},\bxi_{\sth}\} }
  (-\beta_1)^{1-l_{\so}}(-\beta_2)^{1-l_{\st}}
\; f(\bxi_{\so},\bxi_{\st})f(\bxi_{\so},\bxi_{\sth})f(\bxi_{\sth},\bxi_{\st})\\
 \times\lambda_2(\bxi_{\so})\lambda_1(\bxi_{\st}) K^{(1)}_{1,l_{\so}}(u|\bxi_{\so}+c) \overline{K}^{(1)}_{1,l_{\st}}(u|\bxi_{\st}-c)
 \nu_{12}(\{v_m,\bxi_{\sth}\})|0\rangle,
\end{multline}
where $\bxi=\{u,\bv_m\}$. Let $\bw=\{u,\bv\}$. Then equation \eqref{1-contr1} is equivalent to
\begin{multline}\label{1-contr1w}
 \nu_{12}(v_m)\nu_{21}(u)\nu_{12}(\bv_m)|0\rangle=\sum_{\bw\Rightarrow\{\bw_{\so},\bw_{\st},\bw_{\sth}\} }
  (-\beta_1)^{1-l_{\so}}(-\beta_2)^{1-l_{\st}}
\; \frac{f(\bw_{\so},\bw_{\st})f(\bw_{\so},\bw_{\sth})f(\bw_{\sth},\bw_{\st})}{f(\bw_{\so},v_m)f(v_m,\bw_{\st})}\\
 \times\lambda_2(\bw_{\so})\lambda_1(\bw_{\st}) K^{(1)}_{1,l_{\so}}(u|\bw_{\so}+c) \overline{K}^{(1)}_{1,l_{\st}}(u|\bw_{\st}-c)
 \nu_{12}(\bw_{\sth})|0\rangle.
\end{multline}
Indeed, due to the factor $\bigl(f(\bw_{\so},v_m)f(v_m,\bw_{\st})\bigr)^{-1}$ we have $v_m\notin\bw_{\so}$ and $v_m\notin\bw_{\st}$,
because otherwise the corresponding contribution vanishes. Thus, $v_m\in\bw_{\sth}$. Setting
$\bw_{\so}=\bxi_{\so}$, $\bw_{\st}=\bxi_{\st}$, and $\bw_{\sth}=\{v_m,\bxi_{\sth}\}$ in \eqref{1-contr1w} we immediately
arrive at \eqref{1-contr1}.

The action of the terms $\nu_{11}(v_m)\nu_{22}(u)$ and $\nu_{11}(u)\nu_{22}(v_m)$ in \eqref{sing-com} can be computed using Proposition \ref{mulii}. We omit simple but rather exhausting intermediate calculations and give the
final result:
\begin{multline}\label{com112212-2}
g(v_m,u)\bigl(\nu_{11}(u)\nu_{22}(v_m)-\nu_{11}(v_m)\nu_{22}(u)\bigr)\nu(\bv_m)|0\rangle=
\sum_{\bw\Rightarrow\{\bw_{\so},\bw_{\st},\bw_{\sth}\} }
  (-\beta_1)^{1-l_{\so}}(-\beta_2)^{1-l_{\st}}
\; \\
 \times
 f(\bw_{\so},\bw_{\st})f(\bw_{\so},\bw_{\sth})f(\bw_{\sth},\bw_{\st})
 K^{(1)}_{1,l_{\so}}(u|\bw_{\so}+c) \overline{K}^{(1)}_{1,l_{\st}}(u|\bw_{\st}-c)\\
\times\left(1-\frac1{f(\bw_{\so},v_m)f(v_m,\bw_{\st})}\right)\lambda_2(\bw_{\so})\lambda_1(\bw_{\st}) \nu_{12}(\bw_{\sth})|0\rangle.
\end{multline}
Combining equations \eqref{1-contr1w} and \eqref{com112212-2} we obtain \eqref{nvac2112} for $\#\bv=m$.
Thus, the first step of induction is completed.

Let now { assume that} \eqref{nvac2112} holds for some $n-1$. We prove that then it holds for $\#\bu=n$. The
proof is very similar to the one of proposition~\ref{mulii}, however, it is more bulky.

We act successively { as} $\nu_{21}(\bu)=\nu_{21}(u_n)\nu_{21}(\bu_n)$. Then
\begin{multline}\label{mact1221e-1}
\nu_{21}(\bu)\nu_{12}(\bv)|0\rangle=\nu_{21}(u_n) \sum_{\bxi\Rightarrow\{\bxi_{\so},\bxi_{\st},\bxi_{\sth}\} }
  (-\beta_1)^{n-1-l_{\so}}(-\beta_2)^{n-1-l_{\st}}
\lambda_2(\bxi_{\so})\lambda_1(\bxi_{\st})
\\
\times f(\bxi_{\so},\bxi_{\st})f(\bxi_{\so},\bxi_{\sth})f(\bxi_{\sth},\bxi_{\st})
K^{(1)}_{n-1,l_{\so}}(\bu_n|\bxi_{\so}+c) \overline{K}^{(1)}_{n-1,l_{\st}}(\bu_n|\bxi_{\st}-c)
\nu_{12}(\bxi_{\sth})|0\rangle.
\end{multline}
Here $\bxi=\{\bu_n,\bv\}$, $l_{\so}=\#\bxi_{\so}$, and $l_{\st}=\#\bxi_{\st}$. The action of $\nu_{21}(u_n)$ gives us  { an}  additional sum over partitions
\begin{multline}\label{mact1221e-2}
\nu_{21}(\bu)\nu_{12}(\bv)|0\rangle=\sum_{\bxi\Rightarrow\{\bxi_{\so},\bxi_{\st},\bxi_{\sth}\} }
  (-\beta_1)^{n-l_{\so}}(-\beta_2)^{n-l_{\st}}
\lambda_2(\bxi_{\so})\lambda_1(\bxi_{\st})
\; f(\bxi_{\so},\bxi_{\st})f(\bxi_{\so},\bxi_{\sth})f(\bxi_{\sth},\bxi_{\st})\\
\times K^{(1)}_{n-1,l_{\so}}(\bu_n|\bxi_{\so}+c) \overline{K}^{(1)}_{n-1,l_{\st}}(\bu_n|\bxi_{\st}-c)
\sum_{\bet\Rightarrow\{\bet_{\so},\bet_{\st},\bet_{\sth}\} }
(-\beta_1)^{-k_{\so}}(-\beta_2)^{-k_{\st}} \lambda_2(\bet_{\so})\lambda_1(\bet_{\st})\\
\times
\; f(\bet_{\so},\bet_{\st})f(\bet_{\so},\bet_{\sth})f(\bet_{\sth},\bet_{\st})
K^{(1)}_{1,k_{\so}}(u_n|\bet_{\so}+c) \overline{K}^{(1)}_{1,k_{\st}}(u_n|\bet_{\st}-c)
 \nu_{12}(\bet_{\sth})|0\rangle,
\end{multline}
where $\bet={\{\bxi_{\sth},v_n\}}$, $k_{\so}=\#\bet_{\so}$, and $k_{\st}=\#\bet_{\st}$.
Thus, eventually the sum is taken over partitions of the set $\{\bu,\bv\}$ into five subsets $\bxi_{\so}$,
$\bxi_{\st}$, $\bet_{\so}$, $\bet_{\st}$, and $\bet_{\sth}$ such that
$u_n\notin\{\bxi_{\so},\bxi_{\st}\}$. The subset $\bxi_{\sth}$ should be understood as $\bxi_{\sth}=\{\bet_{\so},\bet_{\st},\bet_{\sth}\}
\setminus \{u_n\}$.

Let $\bw=\{\bu,\bv\}$. We denote $\bxi_{\so}=\bw_{\so}$, $\bxi_{\st}=\bw_{\st}$, $\bet_{\so}=\bw_{\qo}$, $\bet_{\qt}=\bw_{\qt}$,
and $\bet_{\sth}=\bw_{\qth}$. Respectively, the cardinalities of the subsets are denoted by
$r_{\so}=\#\bw_{\so}$, $r_{\st}=\#\bw_{\st}$, $r_{\qo}=\#\bw_{\qo}$, $r_{\qt}=\#\bw_{\qt}$. Then equation \eqref{mact1221e-2} takes the form
\begin{multline}\label{mact1221e-3}
\nu_{21}(\bu)\nu_{12}(\bv)|0\rangle=\sum_{\bw\Rightarrow\{\bw_{\so},\bw_{\st},\bw_{\qo},\bw_{\qt},\bw_{\qth}\}}
(-\beta_1)^{n-r_{\so}-r_{\qo}}(-\beta_2)^{n-r_{\st}-r_{\qt}}
\lambda_2(\bw_{\qo})\lambda_2(\bw_{\so})
\lambda_1(\bw_{\st})\lambda_1(\bw_{\qt})
\\
\times\frac{f(\bw_{\so},\bw_{\st})f(\bw_{\so},\bw_{\qo})f(\bw_{\so},\bw_{\qt}) f(\bw_{\so},\bw_{\qth})
f(\bw_{\qo},\bw_{\st})f(\bw_{\qt},\bw_{\st})f(\bw_{\qth},\bw_{\st})}
{f(\bw_{\so},u_n)f(u_n,\bw_{\st})}
\; f(\bw_{\qo},\bw_{\qt})f(\bw_{\qo},\bw_{\qth})f(\bw_{\qth},\bw_{\qt})\\
\times K^{(1)}_{n-1,r_{\so}}(\bu_n|\bw_{\so}+c) K^{(1)}_{1,r_{\qo}}(u_n|\bw_{\qo}+c)
\overline{K}^{(1)}_{n-1,r_{\st}}(\bu_n|\bw_{\st}-c)\overline{K}^{(1)}_{1,r_{\qt}}(u_n|\bw_{\qt}-c)\nu_{12}(\bw_{\qth})|0\rangle.
\end{multline}
Observe that the restriction $u_n\notin\{\bw_{\so},\bw_{\st}\}$ holds automatically, because $\bigl(f(\bw_{\so},u_n)f(u_n,\bw_{\st})\bigr)^{-1}=0$
{ for} $u_n\in\{\bw_{\so},\bw_{\st}\}$.
Setting $\{\bw_{\st},\bw_{\qt}\}=\bw_{0}$, $\{\bw_{\so},\bw_{\qo}\}=\bw_{0'}$,  and $\bw_{\qth}=\bw_{\sth}$ we recast
\eqref{mact1221e-3} as follows:
\begin{multline}\label{mact1221e-4}
\nu_{21}(\bu)\nu_{12}(\bv)|0\rangle=\sum_{\bw=\{\bw_{0'},\bw_{0},\bw_{\sth}\}} (-\beta_1)^{n-r_{0'}}(-\beta_2)^{n-r_{0}}\lambda_2(\bw_{0'})\lambda_1(\bw_{0})\nu_{12}(\bw_{\sth})|0\rangle\\
\times f(\bw_{0'},\bw_{0})f(\bw_{0'},\bw_{\sth})f(\bw_{\sth},\bw_{0})\; W\overline{W},
\end{multline}
where $r_{0'}=\#\bw_{0'}$, $r_{0}=\#\bw_{0}$, and
\be{W}
W=\sum_{\bw_{0'}\Rightarrow\{\bw_{\so},\bw_{\qo}\}}\frac{f(\bw_{\so},\bw_{\qo})}{f(\bw_{\so},u_n)}
K^{(1)}_{n-1,r_{\so}}(\bu_n|\bw_{\so}+c) K^{(1)}_{1,r_{\qo}}(u_n|\bw_{\qo}+c),
\ee
\be{oW}
\overline{W}=\sum_{\bw_{0}\Rightarrow\{\bw_{\st},\bw_{\qt}\}}
\frac{f(\bw_{\qt},\bw_{\st})}{f(u_n,\bw_{\st})}
\overline{K}^{(1)}_{n-1,r_{\st}}(\bu_n|\bw_{\st}-c)\overline{K}^{(1)}_{1,r_{\qt}}(u_n|\bw_{\qt}-c).
\ee

Observe that the sums over partitions of the subsets $\bw_{0'}$ and $\bw_{0}$ can be obtained one from another
via the replacement $c\to -c$. Moreover, the sum \eqref{oW} was computed in \eqref{UseCML-2}. Thus,
\be{WoW}
W= K^{(1)}_{n,r_{0'}}(\bu|\bw_{0'}+c),\qquad \overline{W}=\overline{K}^{(1)}_{n,r_{0}}(\bu|\bw_{0}-c).
\ee
Substituting this into \eqref{mact1221e-4} { we obtain}
\begin{multline}\label{mact1221e-5}
\nu_{21}(\bu)\nu_{12}(\bv)|0\rangle=\sum_{\bw=\{\bw_{0'},\bw_{0},\bw_{\sth}\}} (-\beta_1)^{n-r_{0'}}(-\beta_2)^{n-r_{0}}\lambda_2(\bw_{0'})\lambda_1(\bw_{0})\nu_{12}(\bw_{\sth})|0\rangle\\
\times f(\bw_{0'},\bw_{0})f(\bw_{0'},\bw_{\sth})f(\bw_{\sth},\bw_{0})\; K^{(1)}_{n,r_{0'}}(\bu|\bw_{0'}+c)
\overline{K}^{(1)}_{n,r_{0}}(\bu|\bw_{0}-c),
\end{multline}
{ which} coincides with \eqref{nvac2112} up to the labels of the subsets. This ends the proof.
\endproof

\subsection{Multiple action of the modified operator $\nu_{12}$\label{Mul-Ac-BV12}}

Up to now all the multiple action formulas were valid for an arbitrary highest wight representation of
the Yangian of $\mathfrak{gl}_2$. The following proposition is valid for finite dimensional representations only.

\begin{prop}\label{mul12}
{ Let $\#u=n$ and $\#v=m$. Consider an irreducible finite dimensional representation of the Yangian.
Then there exists an integer $S$ and a function $F(u)$ such that for all $n$ and $m$ such that $m+n \geq S$
the following multiple action holds:}
\begin{multline}\label{mact1212e}
\nu_{12}(\bu)\nu_{12}(\bv)|0\rangle=\nu_{12}(\bw)|0\rangle=\Big(\frac{(\mu-1)(\beta_1+\beta_2)}{\beta_1\beta_2}\Big)^{m+n-S}\\
\times\sum_{\bw\Rightarrow\{\bw_{\so},\bw_{\st}\}} F(\bw_{\so})g(\bw_{\so},\bw_{\st})\nu_{12}(\bw_{\st})|0\rangle.
\end{multline}
The sum is taken over partitions $\{\bv,\bu\}=\bw\Rightarrow\{\bw_{\so},\bw_{\st}\}$ such that
$\#\bw_{\so}=m+n-S$, $\#\bw_{\st}=S$. The constant $\mu$ is defined in  \eqref{Mat-Tf}.
\end{prop}

\begin{rmk}
{ The value of $S$ and the explicit form of the function $F(u)$ depend on the concrete representation \cite{BSV18}.
In particular, for the} case of the fundamental representation of the inhomogeneous  XXX spin-$1/2$ chain with $N$ sites one has
$S=N$ and
\be{F-XXX}
F(u)=\prod_{i=1}^N\frac{h(u,\theta_i)}{g(u,\theta_i)},
\ee
where $\theta_i$ are inhomogeneity parameters.
\end{rmk}
\begin{rmk}
{ Equation \eqref{mact1212e} shows that if the number of the operators $\nu_{12}$ exceeds $S$, then their
successive action on $|0\rangle$ reduces to the action of exactly $S$ such  operators. This property is a peculiarity
of finite-dimensional representations, and it is this property that is key for implementation of the MABA.
In particular, the  function $F(u)$ gives rise to the inhomogeneous term introduced in the context
of the {\it off-diagonal  Bethe ansatz} \cite{CYSW13a,CYSW13c}.}
\end{rmk}
%
\proof
To prove proposition~\ref{mul12} we use induction over $n=\#\bu$ with $n+m \geq S$.
The case $n=1$ was first conjectured in \cite{BP15} for the fundamental representation. Then, it was proved in \cite{BSV18} that for any irreducible finite dimension representation there exists an integer $S$ and a function $F(u)$ such that
\begin{multline} \label{act1212e}
\nu_{12}(u)\nu_{12}(\bv)=\frac{(\mu-1)(\beta_1+\beta_2)}{\beta_1\beta_2} \Big(F(u)g(u,\bv)\nu_{12}(\bv)\\
+\sum_{i=1}^S g(v_i,u)F(v_i)g(v_i,\bv_i)\nu_{12}(u)\nu_{12}(\bv_i)\Big).
\end{multline}
The reader can find  the explicit form of  $F(u)$ and the corresponding $S$ in \cite{BSV18}. { It is easy to see that the term in the first line
of \eqref{act1212e} corresponds to the partition $\bw_{\so}=u$, $\bw_{\st}=\bv$ in \eqref{mact1212e}. The terms in the second line
of \eqref{act1212e} correspond to the partitions $\bw_{\so}=v_i$, $\bw_{\st}=\{u,\bv_i\}$ ($i=1,\dots,m)$ in \eqref{mact1212e}.
Thus,  \eqref{mact1212e} coincides with \eqref{act1212e} for $n=1$.}

Let  \eqref{mact1212e} be valid for $n-1=\#\bu_n$ such that $n-1+m > S$. {Consider the action of $\nu_{12}(\bu)$ with  $n=\#\bu$.
We can act successively,} firstly by $\nu_{12}(\bar u_n)$ and secondly by $\nu_{12}(u_n)$. { Due to the induction assumption we obtain at the first step}
\ben
\nu_{12}(\bar u)\nu_{12}(\bv)|0\rangle=\Big(\frac{(\mu-1)(\beta_1+\beta_2)}{\beta_1\beta_2}\Big)^{m+n-S-1}
\sum_{\bxi \Rightarrow \{\bxi_{\so},\bxi_{\st}\}} F(\bxi_{\so})g(\bxi_{\so},\bxi_{\st})\nu_{12}(u_n)\nu_{12}(\bxi_{\st})|0\rangle,
\een
where the sum is taken over partitions $\bxi=\{\bu_n,\bv\} \Rightarrow \{\bxi_{\so},\bxi_{\st}\}$ such that
$\#\bxi_{\so}=n-1$, $\#\bxi_{\st}=m$. { Acting with $\nu_{12}(u_n)$ on $\nu_{12}(\bxi_{\st})|0\rangle$
via \eqref{act1212e}  we find
\begin{multline}\label{bxibet}
\nu_{12}(\bar u)\nu_{12}(\bv)|0\rangle=\Big(\frac{(\mu-1)(\beta_1+\beta_2)}{\beta_1\beta_2}\Big)^{m+n-S}\\
\times\sum_{\bxi \Rightarrow \{\bxi_{\so},\bxi_{\st}\}}\sum_{\bet \Rightarrow \{\bet_{\so},\bet_{\st}\}}
F(\bxi_{\so})F(\bet_{\so})g(\bxi_{\so},\bxi_{\st})g(\bet_{\so},\bet_{\st})\nu_{12}(\bet_{\st})|0\rangle,
\end{multline}
where we have additional partitions $\bet=\{u_n,\bxi_{\st}\}\Rightarrow  \{\bet_{\so},\bet_{\st}\}$ such that
$\#\bet_{\so}=1$ and $\#\bet_{\st}=m$. Thus, eventually we deal with the partitions of the set $\bw=\{\bu,\bv\}$
into three subsets: $\bxi_{\so}$, $\bet_{\so}$, and $\bet_{\st}$. The subset $\bxi_{\st}$ should be understood as
$\bxi_{\st}=\{\bet_{\so},\bet_{\st}\}\setminus \{u_n\}$. Besides the restrictions on the cardinalities of the subsets
we have { the} additional restriction $u_n\notin\bxi_{\so}$.

Let $\bxi_{\so}=\bw_{\so}$, $\bet_{\so}=\bw_{\st}$, and $\bet_{\st}=\bw_{\sth}$. Then $\bxi_{\st}=\{\bw_{\st},\bw_{\sth}\}\setminus \{u_n\}$,
and equation \eqref{bxibet} takes the form
\begin{multline}\label{bxibet-bw}
\nu_{12}(\bar u)\nu_{12}(\bv)|0\rangle=\Big(\frac{(\mu-1)(\beta_1+\beta_2)}{\beta_1\beta_2}\Big)^{m+n-S}
\sum_{\bw \Rightarrow \{\bw_{\so},\bw_{\st},\bw_{\sth}\}}F(\bw_{\so})F(\bw_{\st})\nu_{12}(\bw_{\sth})|0\rangle\\
\times
\frac{g(\bw_{\so},\bw_{\st})g( \bw_{\so},\bw_{\sth}) }{g(\bw_{\so},u_n)}g(\bw_{\st},\bw_{\sth}).
\end{multline}
Observe that the condition $u_n\notin\bw_{\so}$ is valid automatically due to the factor $\Big(g(\bw_{\so},u_n)\Big)^{-1}$ that vanishes if
$u_n \in \bw_{\so}$.
Setting $\bw_{0}=\{\bw_{\so},\bw_{\st}\}$ we recast \eqref{bxibet-bw} as follows:
\begin{multline}\label{bxibet-bw1}
\nu_{12}(\bar u)\nu_{12}(\bv)|0\rangle=\Big(\frac{(\mu-1)(\beta_1+\beta_2)}{\beta_1\beta_2}\Big)^{m+n-S}
\sum_{\bw \Rightarrow \{\bw_{0},\bw_{\sth}\}}F(\bw_{0})g(\bw_{0},\bw_{\sth})\nu_{12}(\bw_{\sth})|0\rangle\\
\times \sum_{\bw_{0} \Rightarrow \{\bw_{\so},\bw_{\st}\}}
\frac{g(\bw_{\so},\bw_{\st})  }{g(\bw_{\so},u_n)}.
\end{multline}
The sum over partitions is now taken { in} two steps. First, the set $\bw=\{\bu,\bv\}$ is divided into subsets $\{\bw_{0},\bw_{\sth}\}$
such that $\#\bw_{0}=n$ and $\#\bw_{\sth}=m$. Then the subset $\bw_{0}$ is divided into subsets $\{\bw_{\so},\bw_{\st}\}$
such that $\#\bw_{\so}=n-1$ and $\#\bw_{\st}=1$.
Let us prove that the latter sum is equal to $1$.  We have
\begin{equation}\label{PR-1}
\sum_{\bw_{0} \Rightarrow \{\bw_{\so},\bw_{\st}\}}
\frac{g(\bw_{\so},\bw_{\st})  }{g(\bw_{\so},u_n)}=\lim_{x\to u_n}
\frac1{g(\bw_{0},x)}\sum_{\bw_{0} \Rightarrow \{\bw_{\so},\bw_{\st}\}}
g(\bw_{\so},\bw_{\st})  g(\bw_{\st},x).
\end{equation}
Here we have replaced $u_n$ by $x$ in order to avoid possible singularity at $\bw_{\st}=u_n$. Recall that $\#\bw_{\st}=1$. Thus, the sum
over partitions in the { right hand side} of \eqref{PR-1} is given by a contour integral
\ben
\sum_{\bw_{0} \Rightarrow \{\bw_{\so},\bw_{\st}\}}
g(\bw_{\so},\bw_{\st})  g(\bw_{\st},x)=\frac{-1}{2 \pi i c}\oint_{\Gamma(\bw_{0})} g(\bw_{0},z)  g(z,x)\,dz,
\een
where anticlockwise oriented contour $\Gamma(\bw_{0})$ surrounds the points $\bw_{0}$ and does not contain any other singularities
of the integrand. Taking the integral by the residue outside the integration contour (that is, at $z=x$) we immediately obtain
\ben
\sum_{\bw_{0} \Rightarrow \{\bw_{\so},\bw_{\st}\}}
g(\bw_{\so},\bw_{\st})  g(\bw_{\st},x)=g(\bw_{0},x),
\een
leading to
\begin{equation}\label{PR-2}
\sum_{\bw_{0} \Rightarrow \{\bw_{\so},\bw_{\st}\}}
\frac{g(\bw_{\so},\bw_{\st})  }{g(\bw_{\so},u_n)}=1.
\end{equation}
Substituting this into \eqref{bxibet-bw1} we arrive at
\begin{equation}\label{bxibet-bw2}
\nu_{12}(\bar u)\nu_{12}(\bv)|0\rangle=\Big(\frac{(\mu-1)(\beta_1+\beta_2)}{\beta_1\beta_2}\Big)^{m+n-S}
\sum_{\bw \Rightarrow \{\bw_{0},\bw_{\sth}\}}F(\bw_{0})g(\bw_{0},\bw_{\sth})\nu_{12}(\bw_{\sth})|0\rangle,
\end{equation}
{ which} coincides with \eqref{mact1212e} up to the labels of the subsets.
Thus, the proof is completed.

\endproof



\section{Modified scalar product}\label{MSP}

We can now consider the scalar product of the modified Bethe vectors.

\begin{thm}\label{MSPfor}
Let $\#\bu=n$ and $\#\bv=m$. Then the scalar product of two modified Bethe vectors
\be{SPDEF}
S_\nu^{n,m}(\bu,\bv)=\lvec \nu_{21}(\bu)\nu_{12}(\bv)\rvec
\ee
is given by
\begin{equation}\label{SP-fin}
S_\nu^{n,m}(\bu,\bv)=\sum_{\bxi\Rightarrow\{\bxi_{\so},\bxi_{\st}\}}
(-\beta_1)^{n-l_{\so}}(-\beta_2)^{n-l_{\st}}
 \lambda_2(\bxi_{\so})\lambda_1(\bxi_{\st})
f(\bxi_{\so},\bxi_{\st})K_{n,l_{\so}}^{(\mu)}(\bu|\bxi_{\so}+c) \overline{K}^{(\mu)}_{n,l_{\st}}(\bu|\bxi_{\st}-c).
\end{equation}
Here $\bxi=\{\bu,\bv\}$, $l_{\so}=\bxi_{\so}$, and $l_{\st}=\bxi_{\st}$. The sum is taken over all partitions
$\bxi\Rightarrow\{\bxi_{\so},\bxi_{\st}\}$. There is no restriction on the cardinalities of the subsets.
The functions $K^{(\mu)}_{n,l_{\so}}$ and $\overline{K}^{(\mu)}_{n,l_{\st}}$  respectively are the modified
Izergin determinants \eqref{defKdef1} and \eqref{CdefKdef1} at $z=\mu$.

\end{thm}
%
\proof
Acting with the dual highest weight vector \eqref{dHWRG} onto  \eqref{nvac2112}  we find
\begin{multline}\label{SP-init}
S_\nu^{n,m}(\bu,\bv)=\sum_{\bxi\Rightarrow\{\bxi_{\so},\bxi_{\st},\bxi_{\sth}\}}
(-\beta_1)^{n-l_{\so}}(-\beta_2)^{n-l_{\st}}\lambda_2(\bxi_{\so})\lambda_1(\bxi_{\st})
\; f(\bxi_{\so},\bxi_{\st})f(\bxi_{\so},\bxi_{\sth})f(\bxi_{\sth},\bxi_{\st})\\
\times K^{(1)}_{n,l_{\so}}(\bu|\bxi_{\so}+c)\overline{K}^{(1)}_{n,l_{\st}}(\bu|\bxi_{\st}-c)\langle0|\nu_{12}(\bxi_{\sth})|0\rangle.
\end{multline}
Recall that here $\bxi=\{\bu,\bv\}$, $\#\bxi_{\so}=l_{\so}$, and $\#\bxi_{\st}=l_{\st}$.
The sum is taken over all partitions $\bxi\Rightarrow\{\bxi_{\so},\bxi_{\st},\bxi_{\sth}\}$.

The vacuum average $\langle0|\nu(\bxi_{\sth})|0\rangle$ { was computed} in \cite{BS18}:
\begin{equation}\label{Aver12-0}
\langle0|\nu_{12}(\bw)|0\rangle=(1-\mu)^{p}\sum_{\bw\Rightarrow\{\bw_{\so},\bw_{\st}\}} (-\beta_2)^{-\#\bw_{\st}}(-\beta_1)^{-\#\bw_{\so}} \lambda_2(\bw_{\so})\lambda_1(\bw_{\st})\; f(\bw_{\so},\bw_{\st}),
\end{equation}
where $\#\bw=p$ { and the sum is taken over all partitions $\bw\Rightarrow\{\bw_{\so},\bw_{\st}\}$}. Substituting \eqref{Aver12-0} into \eqref{SP-init}  and decomposing  $\bxi_{\sth}=\{\bxi_{\qo},\bxi_{\qt}\}$ we find
\begin{multline}\label{SP-2}
S_\nu^{n,m}(\bu,\bv)=\sum_{\bxi\Rightarrow\{\bxi_{\so},\bxi_{\st},\bxi_{\qo}, \bxi_{\qt}\} }
(1-\mu)^{l_{\qo}+l_{\qt}}
(-\beta_1)^{n-l_{\so}-l_{\qo}}(-\beta_2)^{n-l_{\st}-l_{\qt}}
\lambda_2(\bxi_{\so})\lambda_1(\bxi_{\st})
\lambda_2(\bxi_{\qo})\lambda_1(\bxi_{\qt})\\
\times\; f(\bxi_{\so},\bxi_{\st})f(\bxi_{\so},\bxi_{\qo}) f(\bxi_{\so},\bxi_{\qt})f(\bxi_{\qo},\bxi_{\st})f(\bxi_{\qt},\bxi_{\st})f(\bxi_{\qo},\bxi_{\qt})
K^{(1)}_{n,l_{\so}}(\bu|\bxi_{\so}+c)\overline{K}^{(1)}_{n,l_{\st}}(\bu|\bxi_{\st}-c).
\end{multline}
Here the sum is taken over partitions $\bxi\Rightarrow\{\bxi_{\so},\bxi_{\st},\bxi_{\qo},\bxi_{\qt}\}$. {The cardinalities
of the subsets are denoted by $l$ with the corresponding subscript.}

Now we set $\{\bxi_{\so},\bxi_{\qo}\}=\bxi_{0}$, $\{\bxi_{\st},\bxi_{\qt}\}=\bxi_{0'}$. Then we arrive at {%
\begin{equation}\label{SP-3}
S_\nu^{n,m}(\bu,\bv)=\sum_{\bxi\Rightarrow\{\bxi_{0},\bxi_{0'}\}}
(-\beta_1)^{n-l_{0}}(-\beta_2)^{n-l_{0'}}
\lambda_2(\bxi_{0})\lambda_1(\bxi_{0'})
f(\bxi_{0},\bxi_{0'})\mathcal{L}(\bxi_{0})\overline{\mathcal{L}}(\bxi_{0'}),
\end{equation}
where
\be{cL}
\mathcal{L}(\bxi_{0})= \sum_{\bxi_{0}\Rightarrow\{\bxi_{\so},\bxi_{\qo}\}}
(1-\mu)^{l_{\qo}} K^{(1)}_{n,l_{\so}}(\bu|\bxi_{\so}+c) f(\bxi_{\so},\bxi_{\qo})
\ee
and
\be{ocL}
\overline{\mathcal{L}}(\bxi_{0'})=
\sum_{\bxi_{0'}\Rightarrow\{\bxi_{\st},\bxi_{\qt}\}}
(1-\mu)^{l_{\qt}}\overline{K}^{(1)}_{n,l_{\st}}(\bu|\bxi_{\st}-c)f(\bxi_{\qt},\bxi_{\st}).
\ee
The sums \eqref{cL} and \eqref{ocL} are computed in proposition~\ref{SUM-Kf}:
\be{cL-3}
\mathcal{L}(\bxi_{0})= K^{(\mu)}_{n,l_{0}}(\bu|\bxi_{0}+c), \qquad
\overline{\mathcal{L}}(\bxi_{0'})= \overline{K}^{(\mu)}_{n,l_{0'}}(\bu|\bxi_{0'}-c).
\ee
Then equation \eqref{SP-3} coincides with \eqref{SP-fin} up to the labels of the subsets.
}

\endproof

Remarkably, this formula has exactly the same form as representation \eqref{SCe} for the scalar product in the usual ABA (for $m=n$). However,
instead of the ordinary Izergin determinants we have now modified Izergin determinants. Furthermore, we have
no restrictions on the cardinalities of the subsets.

Consider the case $\mu=1$ and $n=m$. Then, due to \eqref{K10} a non-vanishing contribution occurs if and only if $n\ge\#\bxi_{\so}$ and
$n\ge\#\bxi_{\st}$. Since $\#\bxi_{\so}+\#\bxi_{\st}=2n$, we conclude that
$n=\#\bxi_{\so}$ and $n=\#\bxi_{\st}$. This  leads us to
\begin{equation}\label{SP-fin00}
S_\nu^{n,n}(\bu,\bv)\Bigr|_{\mu=1}=\sum_{\substack{\bxi\Rightarrow\{\bxi_{\so},\bxi_{\st}\}
\\ \#\bxi_{\so}=\#\bxi_{\st}=n}}
 \lambda_2(\bxi_{\so})\lambda_1(\bxi_{\st})
f(\bxi_{\so},\bxi_{\st})K_{n}(\bu|\bxi_{\so}+c) \overline{K}_{n}(\bu|\bxi_{\st}-c),
\end{equation}
and we reproduce the usual ABA scalar product $S^n_t$ given by theorem~\ref{SC}.

Similarly to \eqref{SCbe} the sum \eqref{SP-fin} can be written in the form of  { the}  sum over independent partitions of the sets $\bu$ and $\bv$
({\it modified Izergin--Korepin formula}).
\begin{cor}
Let $\#\bu=n$ and $\#\bv=m$. Then the modified scalar product of two Bethe vectors is given by
\begin{multline}\label{SP-fin-IK}
S_\nu^{n,m}(\bu,\bv)= \mu^{2n}(1-\mu)^{m-n}\sum  (-\beta_1)^{n_2-m_2}(-\beta_2)^{n_1-m_1}
\lambda_2(\bu_{\so}) \lambda_2(\bv_{\st}) \lambda_1(\bu_{\st}) \lambda_1(\bv_{\so})\\
 \times f(\bu_{\so},\bu_{\st})f(\bv_{\st},\bv_{\so})K_{m_2,n_{2}}^{(1/\mu)}(\bv_{\st}|\bu_{\st}) \overline{K}^{(1/\mu)}_{m_1,n_{1}}(\bv_{\so}|\bu_{\so}),
\end{multline}
where the sum is taken over all partitions $\bu \Rightarrow \{\bu_{\so},\bu_{\st}\}$ and $ \bv \Rightarrow \{\bv_{\so},\bv_{\st}\}$ such that $\#\bv_{\so}=m_{1}$, $\#\bv_{\st}=m_{2}$ and $\#\bu_{\so}=n_{1}$, $\#\bu_{\st}=n_{2}$, where $n_1=0,1,\dots,n$ and $m_1=0,1,\dots,m$ .
\end{cor}
\proof
We set $\bw_{\so}\Rightarrow \{\bu_{\so},\bv_{\st}\}$ and $\bw_{\st}\Rightarrow \{\bu_{\st},\bv_{\so}\}$ with $\#\bu_{\so}=n_1$, $\#\bv_{\so}=m_1$, $\#\bu_{\st}=n_2$, $ \#\bv_{\st}=m_2$ and $n=n_1+n_2$,  $m=m_1+m_2$ in \eqref{SP-fin}. Using \eqref{Kz} and \eqref{oKz} we obtain:
\begin{multline}\label{00SCbe}
S_\nu^{n,m}(\bu,\bv)=(-\mu)^n\sum_{\substack{\bu \Rightarrow \{\bu_{\so},\bu_{\st}\}\\ \bv \Rightarrow \{\bv_{\so},\bv_{\st}\} }} (-\beta_1)^{n-n_{1}-m_2}(-\beta_2)^{n-n_2-m_1}
\lambda_2(\bu_{\so}) \lambda_2(\bv_{\st}) \lambda_1(\bu_{\st}) \lambda_1(\bv_{\so})
\\
\times K^{(\mu)}_{n_2,m_2}(\bu_{\st}|\bv_{\st}+c)\overline{K}^{(\mu)}_{n_1,m_1}(\bu_{\so}|\bv_{\so}-c)  f(\bu_{\so},\bu_{\st})f(\bv_{\st},\bv_{\so})f(\bu_{\so},\bv_{\so})f(\bv_{\st},\bu_{\st}).
\end{multline}
Then the use of   \eqref{Kinv1} and  \eqref{cKinv1}
for the modified Izergin determinants immediately gives  \eqref{SP-fin-IK}.
\endproof

{%
\section*{Conclusion\label{S-CON}}

We considered multiple actions of the modified monodromy matrix entries on the modified Bethe vectors within the
framework of the MABA. We shown that they look very similar to the standard multiple actions obtained for the
ordinary ABA in \cite{BPRS12b}. The main difference is that the ordinary Izergin determinant \cite{Ize87} is modified
according to \eqref{defKdef1} { and} \eqref{CdefKdef1}, and the sum over partitions of the Bethe parameters should be taken without restrictions
on the cardinalities of the subsets. The same changes apply to the formula for the scalar product of the modified Bethe vectors.
It would be interesting to compare this result with  those that follow from the separation of variable approach \cite{KMNT15}.

{ Further development of the method, as the one proposed in this paper, can be carried out in several directions. } It is quite
possible that the multiple action formulas { admit} a deformation to  the XXZ model. In this case, however, the property
\eqref{twist-inv} is no longer valid for arbitrary twist matrices. Therefore, one should consider a more
sophisticated face-vertex transformation  of the twist (see {\it e.g.} \cite{ZLCYSW15b,BP152} and references therein).

It is also interesting  to consider models with higher rank algebra. The main open problem in this direction is
to construct the Bethe vectors in the twisted periodic case.

Finally, a very attractive way { for} further development is to consider particular cases of the scalar products of the 
modified Bethe vectors.  It is well known from the ABA
that if one of the vectors is an eigenvector of the transfer matrix (on-shell Bethe vector), then the scalar product admit a
compact determinant representation, which involves the Jacobian of the transfer matrix eigenvalue \cite{Sla89}.
It was conjectured in \cite{BP15} that a similar representation also exists in the case of the scalar products involving the
modified on-shell Bethe vectors. We will provide a proof of this conjecture in our forthcoming publication.
}

\section*{Acknowledgements}
N. S. would like to thank the hospitality of the Institute de Physique Th\'{e}orique at
the CEA de Saclay where a part of this work was done. S.B. and B.V. would like to thank the hospitality of the LMPT of Tours where a part of this work was done. S.B. was supported by a public grant as part of the
Investissement d'avenir project, reference ANR-11-LABX-0056-LMH, LabEx LMH. N.S. was supported by the
Russian Foundation RFBR-18-01-00273a.

\appendix


\section{Properties of  modified Izergin determinant \label{A-PMID}}

In this section we give a list of properties of the modified Izergin determinant introduced in
section~\ref{SS-MID}. In all the propositions listed below $\bu$ and $\bv$ are two sets of arbitrary complex numbers
with cardinalities $\#\bu=n$ and $\#\bv=m$.

\subsection{Basis properties}

\begin{prop}
\be{shiftc}
\begin{aligned}
&K^{(z)}_{n,m}(\bu-c|\bv)=K^{(z)}_{n,m}(\bu|\bv+c),\\
&\overline{K}^{(z)}_{n,m}(\bu-c|\bv)=\overline{K}^{(z)}_{n,m}(\bu|\bv+c).
\end{aligned}
\ee
\be{u-uv-v}
K^{(z)}_{n,m}(-\bu|-\bv)=\overline{K}^{(z)}_{n,m}(\bu|\bv).
\ee
\end{prop}
\proof
These formulas directly follow from \eqref{defKdef1}--\eqref{CdefKdef2} and the definition of the rational functions \eqref{gfh}.
\endproof


\begin{prop}
\be{K0}
K_{n,0}^{(z)}(\bu|\emptyset)=\overline{K}_{n,0}^{(z)}(\bu|\emptyset)=1, \qquad K_{0,n}^{(z)}(\emptyset|\bv)=\overline{K}_{0,n}^{(z)}(\emptyset|\bv)=(1-z)^n,
\ee
\be{K1}
\begin{aligned}
K_{1,m}^{(z)}(u|\bv)&=(1-z)^{m-1}\bigl(f(u,\bv)-z\bigr),\\
K_{n,1}^{(z)}(\bu|v)&=f(\bu,v)-z,\\
\overline{K}_{1,m}^{(z)}(u|\bv)&=(1-z)^{m-1}\bigl(f(\bv,u)-z\bigr),\\
\overline{K}_{n,1}^{(z)}(\bu|v)&=f(v,\bu)-z.\\
\end{aligned}
\ee
\end{prop}
\proof
These formulas directly follow from \eqref{defKdef1}--\eqref{CdefKdef2}.
\endproof


\begin{prop}
\be{Kz}
K_{n+1,m+1}^{(z)}(\{\bu,w-c\}|\{\bv,w\})=-z K_{n,m}^{(z)}(\bu|\bv).
\ee
\be{oKz}
\overline{K}_{n+1,m+1}^{(z)}(\{\bu,w+c\}|\{\bv,w\})=-z \overline{K}_{n,m}^{(z)}(\bu|\bv).
\ee
\end{prop}
\proof
We use representation \eqref{defKdef1}. We see that
only the term $-z\delta_{m+1,k}$ survives in the last row of the determinant due to $f(w-c,w)=0$. Then we obtain
\begin{multline}\label{Kn1}
K_{n+1,m+1}^{(z)}(\{\bu,w-c\}|\{\bv,w\})=-z\det_m\left(-z\delta_{jk}+\frac{f(\bu,v_j)f(w-c,v_j)f(v_j,\bv_j)f(v_j,w)}{h(v_j,v_k)}\right)\\
=-z\det_m\left(-z\delta_{jk}+\frac{f(\bu,v_j)f(v_j,\bv_j)}{h(v_j,v_k)}\right)
=-z K_{n,m}^{(z)}(\bu|\bv),
\end{multline}
because $f(w-c,v_j)f(v_j,w)=1$ due to \eqref{gfh-prop}. Equation \eqref{oKz} then follows from the replacement $c\to -c$.
\endproof

{
\begin{prop}\label{SOP}
\be{K-sumpart}
\begin{aligned}
&K_{n,m}^{(z)}(\bu|\bv)=\sum_{\bv\Rightarrow\{\bv_{\so},\bv_{\st}\}}(-z)^{\#\bv_{\st}} f(\bu,\bv_{\so})f(\bv_{\so},\bv_{\st}),\\
&\overline{K}_{n,m}^{(z)}(\bu|\bv)=\sum_{\bv\Rightarrow\{\bv_{\so},\bv_{\st}\}}(-z)^{\#\bv_{\st}} f(\bv_{\so},\bu)f(\bv_{\st},\bv_{\so}).
\end{aligned}
\ee
Here the sum is taken over all partitions $\bv\Rightarrow\{\bv_{\so},\bv_{\st}\}$.
\be{oK-sumpart}
\begin{aligned}
&K_{n,m}^{(z)}(\bu|\bv)=(1-z)^{m-n}\sum_{\bu\Rightarrow\{\bu_{\so},\bu_{\st}\}}(-z)^{\#\bu_{\so}} f(\bu_{\st},\bv)f(\bu_{\so},\bu_{\st}),\\
&\overline{K}_{n,m}^{(z)}(\bu|\bv)=(1-z)^{m-n}\sum_{\bu\Rightarrow\{\bu_{\so},\bu_{\st}\}}(-z)^{\#\bu_{\so}} f(\bv,\bu_{\st})f(\bu_{\st},\bu_{\so}).
\end{aligned}
\ee
Here the sum is taken over all partitions $\bu\Rightarrow\{\bu_{\so},\bu_{\st}\}$.
\end{prop}

\proof Expanding the determinant \eqref{defKdef1} over diagonal minors we find
\begin{multline}\label{Exp-det}
\det_n\left(\frac{f(v_j,\bv_j)f(\bu,v_j)}{h(v_j,v_k)}-z\delta_{jk}\right)\\
=(-z)^{n}
+\sum_{s=1}^{n}(-z)^{n-s}\sum_{1\le j_1<\dots<j_s\le n}\left(\prod_{p=1}^s f(v_{j_p},\bv_{j_p})f(\bu,v_{j_p})\right)\det_{s}\frac1{h(v_{j_i},v_{j_k})}.
\end{multline}
The determinant in the { right hand side} is the Cauchy determinant, hence,
\be{Cau}
\det_{s}\frac1{h(v_{j_i},v_{j_k})}=\prod_{\substack{p,q=1\\ p\ne q}}^s\frac1{f(v_{j_p},v_{j_q})}.
\ee
Thus, we obtain
\begin{multline}\label{Exp-det1}
\det_n\left(\frac{f(v_j,\bv_j)f(\bu,v_j)}{h(v_j,v_k)}-z\delta_{jk}\right)\\
=(-z)^{n}+\sum_{s=1}^{n}(-z)^{n-s}\sum_{1\le j_1<\dots<j_s\le n}\left(\prod_{p=1}^s
f(v_{j_p},\bv_{j_p})f(\bu,v_{j_p})\right)
\prod_{\substack{p,q=1\\ p\ne q}}^s\frac1{f(v_{j_p},v_{j_q})}.
\end{multline}
This is exactly the sum over partitions given by the first equation \eqref{K-sumpart}. The second equation \eqref{K-sumpart} then follows
by means { of} the replacement $c\to -c$. Equations \eqref{oK-sumpart}
can be proved exactly in the same manner starting from the representation \eqref{defKdef2}.
}
\endproof


\begin{prop}
\be{c-c}
\overline{K}_{n,m}^{(z)}(\bu|\bv)=(1-z)^{m-n}K_{m,n}^{(z)}(\bv|\bu)
\ee
\end{prop}

 \proof Replacing $\bu\leftrightarrow\bv$ and $n\leftrightarrow m$ in \eqref{oK-sumpart} we obtain
\begin{equation}\label{c-c1}
(1-z)^{m-n}K_{m,n}^{(z)}(\bv|\bu)=\sum_{\bv\Rightarrow\{\bv_{\so},\bv_{\st}\}}(-z)^{\#\bv_{\so}} f(\bv_{\st},\bu)f(\bv_{\so},\bv_{\st}).
\end{equation}
Comparing this expansion with the second equation \eqref{K-sumpart} we see that they coincide up to the labels of the subsets. \endproof


\begin{prop}
\be{Kinv1}
K_{n,m}^{(z)}(\bu|\bv+c)=\frac{(-z)^{n}(1-z)^{m-n} }{f(\bv,\bu)}K_{m,n}^{(1/z)}(\bv|\bu).
\ee
\be{cKinv1}
\overline{K}_{n,m}^{(z)}(\bu|\bv-c)=\frac{(-z)^{n}(1-z)^{m-n} }{f(\bu,\bv)}\overline{K}_{m,n}^{(1/z)}(\bv|\bu).
\ee
\end{prop}
\proof Using \eqref{defKdef1} we obtain
\be{Kinv01}
K_{n,m}^{(z)}(\bu|\bv+c)=\det_m\left(-z\delta_{jk}+\frac{f(v_j,\bv_j)}{f(v_j,\bu)h(v_j,v_k)}\right)
=\frac{(-z)^m}{f(\bv,\bu)}\det_m\left(\delta_{jk}f(v_j,\bu)-\frac1z\frac{f(v_j,\bv_j)}{h(v_j,v_k)}\right).
\ee
On the other hand, using \eqref{defKdef2} for $K_{m,n}^{(1/z)}(\bv|\bu)$ we obtain
\be{Kinv02}
K_{m,n}^{(1/z)}(\bu|\bv)=\left(1-\tfrac1z\right)^{n-m}\det_m\left(\delta_{jk}f(v_j,\bu)-\frac1z\frac{f(v_j,\bv_j)}{h(v_j,v_k)}\right).
\ee
Comparing \eqref{Kinv01} and \eqref{Kinv02} we arrive at \eqref{Kinv1}. Equation \eqref{cKinv1} follows from the replacement $c\to -c$.
\endproof


\begin{prop}
The function $K_{n,m}^{(z)}(\bu|\bv)$ has poles at $u_j=v_k$. The residue at $u_n=v_m$ is given by
\be{resK}
\begin{aligned}
&K_{n,m}^{(z)}(\bu|\bv)\Bigr|_{u_n\to v_m}=g(u_n,v_m)f(\bu_n,u_n)f(v_m,\bv_m)K_{n-1,m-1}^{(z)}(\bu_n|\bv_m)+reg,\\
&\overline{K}_{n,m}^{(z)}(\bu|\bv)\Bigr|_{u_n\to v_m}=g(v_m,u_n)f(u_n,\bu_n)f(\bv_m,v_m)\overline{K}_{n-1,m-1}^{(z)}(\bu_n|\bv_m)+reg,
\end{aligned}
\ee
where $reg$ means regular part.
\end{prop}

\proof It is clear that { the} two equations \eqref{resK} are related by the replacement $c\to -c$. To prove the first equation we use \eqref{defKdef2}. Then for  $u_n=v_m$ the { pole} occurs only in the matrix element
$\delta_{nk}f(u_n,\bv)$. The determinant reduces to the product of this element and the corresponding minor:
\begin{multline}\label{ini-sum1}
K_{n,m}^{(z)}(\bu|\bv)\Bigr|_{u_n\to v_m}=(1-z)^{m-n}g(u_n,v_m)f(v_m,\bv_m)\\
\times\det_{n-1}\left(\delta_{jk}f(u_j,\bv_m)f(u_j,v_m)-z\frac{f(u_j,\bu_{j,n})f(u_j,u_n)}{h(u_j,u_k)}\right)+reg,
\end{multline}
where $\bu_{j,n}=\bu\setminus\{u_j,u_n\}$.  We see that for $u_n=v_m$ we can extract the factor $f(u_j,u_n)$ form the
$j$-th row of the matrix. Thus,
\begin{multline}\label{ini-sum11}
K_{n,m}^{(z)}(\bu|\bv)\Bigr|_{u_n\to v_m}=(1-z)^{m-n}g(u_n,v_m)f(v_m,\bv_m)f(\bu_n,u_n)\\
\times\det_{n-1}\left(\delta_{jk}f(u_j,\bv_m)-z\frac{f(u_j,\bu_{j,n})}{h(u_j,u_k)}\right)+reg,
\end{multline}
{ which} ends the proof.\endproof

\subsection{Summation formulas}

\begin{prop}\label{GenML}
Let $\bxi$, $\bu$, and $\bv$ be sets of arbitrary complex numbers such that $\#\bxi=l$, $\#\bu=n$, and $\#\bv=m$.
Then
\be{ML-1}
\begin{aligned}
&\sum_{\bxi\Rightarrow\{\bxi_{\so},\bxi_{\st}\}}z_2^{l_{\so}}
K_{n,l_{\so}}^{(z_1)}(\bu|\bxi_{\so})K_{m,l_{\st}}^{(z_2)}(\bv|\bxi_{\st})f(\bxi_{\st},\bxi_{\so})f(\bu,\bxi_{\st})
=K_{n+m,l}^{(z_1z_2)}(\{\bu,\bv\}|\bxi),\\
&\sum_{\bxi\Rightarrow\{\bxi_{\so},\bxi_{\st}\}}z_2^{l_{\so}}
\overline{K}_{n,l_{\so}}^{(z_1)}(\bu|\bxi_{\so})\overline{K}_{m,l_{\st}}^{(z_2)}(\bv|\bxi_{\st})f(\bxi_{\so},\bxi_{\st})f(\bxi_{\st},\bu)
=\overline{K}_{n+m,l}^{(z_1z_2)}(\{\bu,\bv\}|\bxi).
\end{aligned}
\ee
Here  $l_{\so}=\#\bxi_{\so}$ and $l_{\st}=\#\bxi_{\st}$. The sums are taken with respect to all partitions $\bxi\Rightarrow\{\bxi_{\so},\bxi_{\st}\}$. There is no restriction on the cardinalities of the subsets.
\end{prop}
\proof
It is clear that { the} two equations \eqref{ML-1} are related by the replacement $c\to -c$. To prove the first equation we use  \eqref{K-sumpart}:
\be{dec}
\begin{aligned}
&K_{n,l_{\so}}^{(z_1)}(\bu|\bxi_{\so})=\sum_{\bxi_{\so}\Rightarrow\{\bxi_{1},\bxi_{2}\}}(-z_1)^{l_{2}} f(\bu,\bxi_{1})f(\bxi_{1},\bxi_{2}),\\
&K_{m,l_{\st}}^{(z_2)}(\bv|\bxi_{\st})=\sum_{\bxi_{\st}\Rightarrow\{\bxi_{3},\bxi_{4}\}}(-z_2)^{l_{4}} f(\bv,\bxi_{3})f(\bxi_{3},\bxi_{4}).
\end{aligned}
\ee
Here we use Arabic numbers for numeration the subsets. The corresponding cardinalities are
$l_i=\#\bxi_i$, $i=1,2,3,4$. Thus, $l_{\so}=l_{1}+l_{2}$ and $l_{\st}=l_{3}+l_{4}$. Denoting the { left hand side} of the first equation \eqref{ML-1} by $\Lambda$ we obtain
\begin{multline}\label{Lam}
\Lambda=\sum_{\bxi\Rightarrow\{\bxi_{1},\bxi_{2},\bxi_{3},\bxi_{4}\}}(-1)^{l_4}z_2^{l_{1}+l_{2}+l_4}(-z_1)^{l_{2}}
f(\bu,\bxi_{1})f(\bv,\bxi_{3})f(\bxi_{1},\bxi_{2})f(\bxi_{3},\bxi_{4})\\
\times
f(\bu,\bxi_{3})f(\bu,\bxi_{4})f(\bxi_{3},\bxi_{1})f(\bxi_{3},\bxi_{2})f(\bxi_{4},\bxi_{1})f(\bxi_{4},\bxi_{2}).
\end{multline}
Setting $\{\bxi_{4},\bxi_{1}\}=\bxi_0$ we find
\begin{multline}\label{Lam1}
\Lambda=f(\bu,\bxi)\sum_{\bxi\Rightarrow\{\bxi_{0},\bxi_{2},\bxi_{3}\}}z_2^{l_0+l_{2}}(-z_1)^{l_{2}}
\frac{f(\bv,\bxi_{3})}{f(\bu,\bxi_{2})}f(\bxi_{0},\bxi_{2})f(\bxi_{3},\bxi_{0})f(\bxi_{3},\bxi_{2})\\
\times \sum_{\bxi_0\Rightarrow\{\bxi_{1},\bxi_{4}\}}
(-1)^{l_4} f(\bxi_{4},\bxi_{1}).
\end{multline}
It was proved in \cite{BS18} that for any set of variables $\bar x$ such that $\#\bar x=p$ the following identity holds:
\be{sum-binom}
\sum_{\substack{\bar x\Rightarrow\{\bar x_{\so},\bar x_{\st}\}\\ \#\bar x_{\so}=k}}
f(\bar x_{\st},\bar x_{\so})=\sum_{\substack{\bar x\Rightarrow\{\bar x_{\so},\bar x_{\st}\}\\ \#\bar x_{\so}=k}}
f(\bar x_{\so},\bar x_{\st})=
\binom{p}{k}.
\ee
Here the sum is taken over partitions $\bar x\Rightarrow\{\bar x_{\so},\bar x_{\st}\}$ such that the cardinality of the subset
$\bar x_{\so}$ is fixed by $\#\bar x_{\so}=k$, $k\le p$. Applying this result to the sum
over partitions  $\bxi_0\Rightarrow\{\bxi_{1},\bxi_{4}\}$  we see that this sum vanishes if $\bxi_0\ne\emptyset$:
\be{sum-nul}
\sum_{\bxi_0\Rightarrow\{\bxi_{1},\bxi_{4}\}}
(-1)^{l_4} f(\bxi_{4},\bxi_{1})=\sum_{l_4=0}^{l_0}(-1)^{l_4}\binom{l_0}{l_4}=(1-1)^{l_0}.
\ee
Thus, we obtain
\begin{equation}\label{Lam2}
\Lambda=\sum_{\bxi\Rightarrow\{\bxi_{2},\bxi_{3}\}}(-z_1z_2)^{l_{2}}
f(\bv,\bxi_{3})f(\bu,\bxi_{3})
f(\bxi_{3},\bxi_{2})=K_{n+m,l}^{(z_1z_2)}(\{\bu,\bv\}|\bxi),
\end{equation}
due to \eqref{K-sumpart}.

Replacing $\bxi$ by $\bxi\pm c$ and setting $z_1=z_2=1$ in \eqref{ML-1} we obtain

\be{CML-2}
\begin{aligned}
&\sum_{\bxi\Rightarrow\{\bxi_{\so},\bxi_{\st}\}}
K_{n,l_{\so}}^{(1)}(\bu|\bxi_{\so}+c)K_{m,l_{\st}}^{(1)}(\bv|\bxi_{\st}+c)\frac{f(\bxi_{\st},\bxi_{\so})}{f(\bxi_{\st},\bu)}
=K_{n+m,k}^{(1)}(\{\bu,\bv\}|\bxi+c),\\
&\sum_{\bxi\Rightarrow\{\bxi_{\so},\bxi_{\st}\}}
\overline{K}_{n,l_{\so}}^{(1)}(\bu|\bxi_{\so}-c)\overline{K}_{m,l_{\st}}^{(1)}(\bv|\bxi_{\st}-c)\frac{f(\bxi_{\so},\bxi_{\st})}{f(\bu,\bxi_{\st})}
=\overline{K}_{n+m,k}^{(1)}(\{\bu,\bv\}|\bxi-c).
\end{aligned}
\ee
These formulas were used in sections~\ref{Mul-Ac-Vac1} and~\ref{Mul-Ac-BV}.

\endproof

\begin{prop}\label{SUM-Kf}
Let $\bu$ and $\bv$ be sets of arbitrary complex numbers such that  $\#\bu=n$ and $\#\bv=m$.
Then
\be{sun-Kf}
\begin{aligned}
&\sum_{\bv\Rightarrow\{\bv_{\so},\bv_{\st}\}}
z_1^{l_{\st}} K^{(z_2)}_{n,l_{\so}}(\bu|\bv_{\so}) f(\bv_{\so},\bv_{\st})
=K_{n,m}^{(z_2-z_1)}(\bu|\bv),\\
&\sum_{\bv\Rightarrow\{\bv_{\so},\bv_{\st}\}}
z_1^{l_{\st}} \overline{K}^{(z_2)}_{n,l_{\so}}(\bu|\bv_{\so}) f(\bv_{\st},\bv_{\so})
=\overline{K}_{n,m}^{(z_2-z_1)}(\bu|\bv).
\end{aligned}
\ee
Here  $l_{\st}=\#\bv_{\st}$. The sums are taken with respect to all partitions $\bv\Rightarrow\{\bv_{\so},\bv_{\st}\}$. There is no restriction on the cardinalities of the subsets.
\end{prop}

\proof
Obviously, the two equations \eqref{sun-Kf} are related by the replacement $c\to -c$, therefore, we prove only the first equation.
Let
\be{sum-L1}
\mathcal{L}= \sum_{\bv\Rightarrow\{\bv_{\so},\bv_{\st}\}}
z_1^{l_{\st}} K^{(z_2)}_{n,l_{\so}}(\bu|\bv_{\so}) f(\bv_{\so},\bv_{\st}).
\ee
Using  \eqref{K-sumpart} we obtain
\be{sum-L2}
\mathcal{L}= \sum_{\bv\Rightarrow\{\bv_{\qo},\bv_{\qt},\bv_{\st}\}}
z_1^{l_{\st}} (-z_2)^{l_{\qt}}  f (\bu,\bv_{\qo})f(\bv_{\qo},\bv_{\qt}) f(\bv_{\qo},\bv_{\st})f(\bv_{\qt},\bv_{\st}).
\ee
Here  $l_{\st}=\#\bv_{\st}$, $l_{\qt}=\#\bv_{\qt}$, and the sum is taken with respect to all partitions $\bv\Rightarrow\{\bv_{\qo},\bv_{\qt},\bv_{\st}\}$.
Setting $\bv_{0}=\{\bv_{\qt},\bv_{\st}\}$ and $l_{0}=\#\bv_{0}$ we find
\be{sum-L3}
\mathcal{L}= \sum_{\bv\Rightarrow\{\bv_{\qo},\bv_{0}\}}
z_1^{l_{0}}   f (\bu,\bv_{\qo})f(\bv_{\qo},\bv_{0})
\sum_{\bv_{0}\Rightarrow\{\bv_{\qt},\bv_{\st}\}}\left(-\tfrac{z_2}{z_1}\right)^{l_{\qt}}f(\bv_{\qt},\bv_{\st}).
\ee
Here we first have the sum over partitions $\bv\Rightarrow\{\bv_{\qo},\bv_{0}\}$ and then the subset $\bv_{0}$ is divided once more as
$\bv_{0}\Rightarrow\{\bv_{\qt},\bv_{\st}\}$. Using \eqref{sum-binom} we find
\be{sum-L4}
\sum_{\bv_{0}\Rightarrow\{\bv_{\qt},\bv_{\st}\}}\left(-\tfrac{z_2}{z_1}\right)^{l_{\qt}}f(\bv_{\qt},\bv_{\st})=
\sum_{l_{\qt}=0}^{l_{0}}\left(-\tfrac{z_2}{z_1}\right)^{l_{\qt}}\binom{l_{0}}{l_{\qt}}
=\left(1-\tfrac{z_2}{z_1}\right)^{l_{0}}.
\ee
Substituting this result into \eqref{sum-L3} we immediately arrive at
\be{sum-L5}
\mathcal{L}= \sum_{\bv\Rightarrow\{\bv_{\qo},\bv_{0}\}}
(z_1-z_2)^{l_{0}}   f (\bu,\bv_{\qo})f(\bv_{\qo},\bv_{0})=K_{n,m}^{(z_2-z_1)}(\bu|\bv),
\ee
due to \eqref{K-sumpart}.\endproof

\section{Commutation relations of  $t_{ij}(u)$ and $\nu_{ij}(u)$}\label{ComY}

The RTT relation \eqref{RTT} yields { to} the following commutation relations:
\be{genCR}
[t_{ij}(u),t_{kl}(v)]=g(u,v)\bigl(t_{kj}(v)t_{il}(u)-t_{kj}(u)t_{il}(v)\bigr).
\ee
In particular,
\ben\label{com1}
&&t_{ij}(u)t_{ij}(v)=t_{ij}(v)t_{ij}(u),\qquad \forall i,j,\\
\label{coms11}
&&t_{11}(u)t_{12}(v)=f(v,u)t_{12}(v)t_{11}(u)+g(u,v)t_{12}(u)t_{11}(v),\\
\label{comsl22}
&&t_{22}(u)t_{12}(v)=f(u,v)t_{12}(v)t_{22}(u)+g(v,u)t_{12}(u)t_{22}(v),\\
\label{comsl2112}
&&[t_{21}(u),t_{12}(v)]=g(u,v)\bigl(t_{11}(v)t_{22}(u)-t_{11}(u)t_{22}(v)\bigr).
\een

In turn, commutation relations \eqref{coms11} { and} \eqref{comsl22} imply the following multiple commutation relations \cite{BPRS12b}:
\be{MCR1122}
\begin{aligned}
&t_{11}(\bu)t_{12}(\bv)= (-1)^n\sum_{\#\bw_{\so}=n} \overline{K}_{n}(\bu|\bw_{\so}+c)f(\bw_{\st},\bw_{\so})t_{12}(\bw_{\st})t_{11}(\bw_{\so}),\\
&t_{22}(\bu)t_{12}(\bv)=(-1)^n\sum_{\#\bw_{\so}=n} K_{n}(\bu|\bw_{\so}+c)f(\bw_{\so},\bw_{\st})t_{12}(\bw_{\st})t_{22}(\bw_{\so}).
\end{aligned}
\ee
Here $\#\bu=n$, $\#\bv=m$,  $\bw=\{\bu,\bv\}$, and $K_{n}$ is the Izergin determinant.
The sums are taken over partitions $\bw \Rightarrow \{\bw_{\so},\bw_{\st}\}$ such that $\#\bw_{\so}=n$.

The same commutation relations are valid for the modified operators $\nu_{ij}(u)$.

\section{Symmetries of the Yangian \label{SymYangian}}

Consider a mapping
\ben\label{auto-mor}
 \phi(T(u))= T^\tau(-u),
\een
where $\tau$ is the diagonal transposition $A^\tau=\sigma^1 A^t \sigma^1$ and $\sigma^1=\big(\begin{smallmatrix}0& 1\\ 1&0\end{smallmatrix}\big)$.
It defines an automorphism of the Yangian of $\mathfrak{gl}_2$ \cite{Mol07}. This automorphism
allows us to find the action of $\nu_{22}(u)$ on the modified Bethe vector knowing { those for} $\nu_{11}(u)$
\ben\label{mor4}
 \phi(\nu_{11}(u)\nu_{12}(\bv)\rvec)=\nu_{22}(-u)\nu_{12}(-\bv)\rvec.
\een
Here we have to apply the following prescriptions: $\phi(\lambda_i(u))=\lambda_{3-i}(-u)$ and $\phi(\beta_i)={\beta_{3-i}}$.

Let us consider as  { an}  example the action  \eqref{nvac1112}.  Applying the mapping $\phi$ to this equation we obtain
\be{Fnvac11-1}
\nu_{22}(-\bu)\nu_{12}(-\bv)|0\rangle =\beta_1^n\sum_{\bw\Rightarrow\{\bw_{\so},\bw_{\st}\}} (-\beta_1)^{-l}
\lambda_2(-\bw_{\so}) \overline{K}^{(1)}_{n,l}(\bu|\bw_{\so}-c)
f(\bw_{\st},\bw_{\so})\nu_{12}(-\bw_{\st})|0\rangle.
\ee
Changing $\bu\to -\bu$ and $\bv\to -\bv$ we arrive at
\be{Fnvac11-2}
\nu_{22}(\bu)\nu_{12}(\bv)|0\rangle =\beta_1^n\sum_{\bw\Rightarrow\{\bw_{\so},\bw_{\st}\}} (-\beta_1)^{-l}
\lambda_2(\bw_{\so}) \overline{K}^{(1)}_{n,l}(-\bu|-\bw_{\so}-c)
f(\bw_{\so},\bw_{\st})\nu_{12}(\bw_{\st})|0\rangle.
\ee
Finally, using \eqref{u-uv-v} we reproduce equation \eqref{nvac2212}.

\end{document}